\documentclass[useAMS,usenatbib,usegraphicx]{mn2e}
\usepackage{amsmath}
\usepackage{amssymb}
\usepackage{multirow}

\def\simlt{\lower.5ex\hbox{$\; \buildrel < \over \sim \;$}}
\def\simgt{\lower.5ex\hbox{$\; \buildrel > \over \sim \;$}}

\newcommand{\bd}{\begin{displaymath}}
\newcommand{\ed}{\end{displaymath}}
\newcommand{\be}{\begin{equation}}
\newcommand{\ee}{\end{equation}}
\newcommand{\beqa}{\begin{eqnarray}}
\newcommand{\eeqa}{\end{eqnarray}}

\newcommand{\Lya}{Ly$\alpha$}

\newcommand{\vbc}{v_{\rm bc}}
\newcommand{\Piso}{P_{\rm iso}}

\title[Rich complexity of 21-cm fluctuations] 
{The rich complexity of 21-cm fluctuations produced by the first stars} 
\author[Fialkov  \& Barkana] {Anastasia Fialkov$^{1}$\thanks{E-mail:
    anastasia.fialkov@phys.ens.fr},
  Rennan Barkana$^{2}$ \\
  $^{1}$ Departement de Physique, Ecole Normale Sup\'{e}rieure, CNRS, 
  24 rue Lhomond, Paris, 75005 France\\
  $^{2}$ Raymond and Beverly Sackler School of Physics and Astronomy,
  Tel Aviv University, Tel Aviv 69978, Israel }

\begin{document}
\pagerange{\pageref{firstpage}--\pageref{lastpage}} \pubyear{2013}
\maketitle

\label{firstpage}

\begin{abstract}
  We explore the complete history of the 21-cm signal in the redshift
  range $z = 7-40$. This redshift range includes various epochs of
  cosmic evolution related to primordial star formation, and should be
  accessible to existing or planned low-frequency radio telescopes. We
  use semi-numerical computational methods to explore the fluctuation
  signal over wavenumbers between 0.03 and 1 Mpc$^{-1}$, accounting for the
  inhomogeneous backgrounds of \Lya, X-ray, Lyman-Werner and ionizing
  radiation. We focus on the recently noted expectation of heating
  dominated by a hard X-ray spectrum from high-mass X-ray binaries. We
  study the resulting delayed cosmic heating and suppression of gas
  temperature fluctuations, allowing for large variations in the 
  minimum halo mass that contributes to star formation. We show that
  the wavenumbers at which the heating peak is detected in
  observations should tell us about the characteristic mean free path
  and spectrum of the emitted photons, thus giving key clues as to the
  character of the sources that heated the primordial Universe. We
  also consider the line-of-sight anisotropy, which allows additional
  information to be extracted from the 21-cm signal. For example, the
  heating transition at which the cosmic gas is heated to the
  temperature of the cosmic microwave background should be clearly
  marked by an especially isotropic power spectrum. More generally, an
  additional cross-power component $P_X$ directly probes which sources
  dominate 21-cm fluctuations. In particular, during cosmic
  reionization (and after the just-mentioned heating transition),
  $P_X$ is negative on scales dominated by ionization fluctuations and
  positive on those dominated by temperature fluctuations.
\end{abstract}


\section{Introduction}

\label{Sec:Intro}

Critical stages of the evolution of our Universe still remain a
mystery due to the lack of direct observational probes. This includes
the cosmic dark ages, the formation era of the first stars and black
holes, the early heating and metal enrichment of the intergalactic
medium (IGM), and the epoch of reionization. A crucial step in the
evolution of the Universe is the formation of first stars. In fact, to
start forming stars in the pristine chemical conditions of the early
Universe, gas has to be gravitationally accelerated at least to the
temperature of $T \simgt 300$ K to initiate the radiative cooling
process (for comparison, atomic hydrogen radiatively cools when it
reaches $T\sim 10^4$ K). Since the cooling temperature of molecular
hydrogen is so low, stars can form even in light halos of mass $\sim
10^5$ M$_\odot$ \citep{Tegmark:1997, Haiman:1996,
  Machacek:2001,Abel:2002}.  The first stars emitted radiation, in
particular photons in the Lyman-Werner (LW) band (11.2-13.6 eV), which
dissociate molecular hydrogen, leading to a suppression of star
formation. As a result of this negative feedback, larger halos were
needed to accelerate gas enough so that it can radiatively cool,
condense and form stars \citep{Machacek:2001, Wise:2007, OShea:2008,
  Haiman:2000}. The effect of a time-dependent LW background is still
highly unconstrained given the current state of observations and
simulations. A first study in a recent numerical simulation
\citep{Visbal:2014} suggests that the effect of LW radiation may be
fast, leading to a relatively strong suppression of H$_2$-based star
formation early in cosmic history. An additional suppression of star
formation via cooling of molecular hydrogen is due to the supersonic
relative velocity ($\vbc$) between dark matter and gas
\citep{Tseliakhovich:2010}.

During its early evolution (before the epoch of reionization) the
Universe remained overall neutral with atomic hydrogen being the most
abundant element. This should allow the probing of most of the as-yet
unobserved stages of cosmic history using lines of hydrogen, in
particular its 21-cm line \citep{Hulst:1945} which corresponds to the
hyper-fine splitting of the H~I ground state. This line, despite its
low optical depth, can in principle be observed today using
ground-based or space and lunar telescopes, allowing us to complete
our knowledge of cosmic history. In fact, many of the low-frequency
radio observatories in place today are after this ``holy grail''. This
includes the Low Frequency Array [LOFAR] \citep{vanHaarlem:2013}, the
Murchison Wide-field Array [MWA] \citep{Bowman:2013}, the Giant Meterwave
Radio Telescope [GMRT] \citep{Paciga:2013}, the Long Wavelength Array
[LWA] \citep{Ellingson:2013}, and the Precision Array to Probe the
Epoch of Reionization [PAPER] \citep{Parsons:2010}. These instruments
are mainly focused on measuring the 21-cm signal from the Epoch of
Reionization around $z\sim10$.  Future observatories, including the
Square Kilometer Array [SKA] \citep{Mellema:2013}, Hydrogen Epoch of
Reionization Array [HERA] \citep{Pober:2014} and the Dark Ages Radio
Explorer [DARE] \citep{Burns:2012} will likely be designed to observe
the redshifted 21-cm signal from higher redshifts. The design of the
next generation experiments as well as interpretation of data from the
telescopes that are currently making observations, rely on our
theoretical understanding of the expected signal.  Modeling of the
21-cm signal from high redshifts is technically very challenging, and
it is a rapidly evolving field.  For instance, the global spectrum and
the power spectrum of the expected 21-cm signal was recently shown to
be very sensitive to the spectral energy distribution (SED) of X-rays
emitted by the first heating sources \citep{Fialkov:2014}.

The origin of the heating sources at high redshifts is still largely
uncertain. High-mass X-ray binaries (HMXBs) are thought to be the most
likely driver of cosmic heating, the idea \citep{Mirabel:2011}
supported by current observations and a recent detailed population
synthesis simulation of X-ray binaries \citep{Fragos:2013} that found
that HMXBs likely dominate over the contribution of quasars at $z
\gtrsim 6-8$. This work was calibrated to all available low-redshift
X-ray observations and yielded X-ray spectra which peak at $\sim 3$
keV and have an almost redshift-independent shape. On the other hand,
the overall normalization of the spectral energy distribution (SED)
changes by an order of magnitude due to the changing metallicity, and
also roughly tracks the evolving cosmic star formation rate.
\citet{Fragos:2013} provides a reasonable framework to study the
effect of X-ray binaries on cosmic history, as in
\citet{Fialkov:2014b}. The probable dominance of HMXBs as the early
heating sources is also (indirectly) supported by numerical
simulations that show that the first stars were massive and likely
formed in groups due to fragmentation (e.g., see the review by
\citet{Bromm:2013} and references therein). Such systems of multiple
heavy stars are likely to evolve into HMXBs.

Other possible heating sources are thermal emission from hot gas in
galaxies \citep{Mineo:2012,Mesinger:2014} and Compton emission from
relativistic electrons \citep{Oh:2001}, which are likely subdominant
at high redshift \citep{Fialkov:2014b}. Specifically, in observed
galaxies the X-ray energy from hot gas is lower than that from X-ray
binaries by a factor of several; at high redshifts, the higher heating
efficiency of soft X-rays (as expected from hot gas) could compensate
for this, but the order of magnitude increase in X-rays from X-ray binaries  at
the low metallicities of early galaxies would have to be compensated
by a similar increase in the hot gas emission.  More promising is the
possibility of X-rays from a population of mini-quasars, i.e., central
black holes in early star-forming halos.  The properties of these
objects are highly unconstrained, so in principle they could produce
either early or late heating \citep{Tanaka:2012, Ciardi:2010};
observationally-based estimates suggest that mini-quasars are also
likely subdominant (see Methods in \citet{Fialkov:2014b}). Heating by
shocks due to structure formation at high redshifts is also unlikely
to be important \citep{shocks}, even after being enhanced by
supersonic relative motion between the gas and dark matter
\citep{McQuinn:2012}. Some more exotic scenarios have also been
suggested, including heating by evaporating black holes
\citep{Ricotti:2008}, cosmic strings \citep{Brandenberger:2010}, or
dark matter annihilation \citep{Valdes:2013}. In any case, our results
also act as an illustration of the more general idea of how the
properties of cosmic heating sources can be probed with 21-cm
observations.

Despite the plethora of possible heating sources, the effect of
inhomogeneous X-rays has not been studied in simulations of
reionization. Due to the historical evolution of the field
\citep{Madau:1997, Furlanetto:2006, Furlanetto:2006b}, many of the
present state-of-the-art simulations of reionization, which include
precise radiative transfer for ionizing photons, neglect the role of
an inhomogeneous X-ray background. For instance, some simulations
(e.g., \citet{Iliev:2014}) assume heating to be saturated during
reionization, i.e., that the gas is much hotter than the Cosmic
Microwave Background (CMB), in which case X-rays do not have an impact
on the 21-cm signal (except in the unlikely case that they contribute
substantially to reionization). Saturated heating is also usually
assumed in analyses of current and future observations of cosmic
reionization \citep{Patil:2014,Pober:2014}. Some simulations do treat
the radiative transfer of X-rays \citep{Baek:2010, Zawada:2014}, but
they do not include X-ray binaries as sources, with one reasonable
argument being that the mean free path of those hard X-rays would be
larger than their simulation box. In addition, most semi-numerical,
hybrid methods, which are capable of making predictions for the 21-cm
signal on very large scales (e.g., \citet{Mesinger:2011, Visbal:2012,
  Fialkov:2013, Fialkov:2014}), used a soft power-law spectrum for
X-ray emission motivated by \citet{Furlanetto:2006b} who used the
results of \citet{Rephaeli:1995}. However, more recent work suggests
that the power-law spectrum does not describe well the observed or
expected X-ray emission by HMXBs, which as mentioned above are
probably the main X-ray sources within starburst galaxies.

Motivated by the necessity to better model the effect of X-rays on the
21-cm signal as discussed above, we estimated the impact of the X-ray
SED on the reionization era ($z = 7-16$) in our previous paper
\citep{Fialkov:2014}. In this work one of our main goals is to extend
the redshift range and to explore the impact of a hard versus soft
X-ray SED on the 21-cm signal from $z = 7-40$, which now includes the
redshifts at which the effect of X-rays is expected to be maximal ($z
\sim 17-20$). A second main point of emphasis in this paper is on the
line-of-sight anisotropy of the 21-cm signal due to the gradient of
the gas velocity. This anisotropy adds some overall power to the 21-cm
fluctuations \citep{Indian}, but more importantly, it allows several
power spectrum components to be separately measured, in principle
allowing much more information to be extracted about the various
sources of 21-cm fluctuations \citep{Barkana:2005a}. We note that the
line-of-sight anisotropy has been previously studied during
reionization
\citep{Mesinger:2011,Mao:2012,Shapiro:2013,Majumdar:2013}, but we
discuss it over a wide redshift range that includes various cosmic
epochs. Finally, we make sure to sample a wide range of parameters.
Specifically, we analyze several cases of possible star-formation
histories, considering three main cases of star-forming halos
including the molecular cooling case with time-dependent LW feedback
and with the velocity streaming effect, the atomic cooling case, and
the third case in which stars form only in massive halos (M$_h\gtrsim
10^8$ M$_\odot$).  
 
This paper is organized as follows: In section~\ref{Sec:Methods} we
briefly discuss our model, including technical aspects of the methods
applied in this paper. We then present and discuss our results for the
cosmic heating history and the 21-cm signal in
section~\ref{Sec:results}. Finally, we summarize our conclusions in
section~\ref{Sec:sum}.

\section{Model and Computational Methods}
\label{Sec:Methods}

In this paper, we use semi-numerical computational methods
\citep{Mesinger:2011,Fialkov:2014b} to efficiently and
self-consistently generate histories of the 21-cm signal in $\sim 400$
Mpc$^3$ volumes, applying a sub-grid model to each 3~Mpc pixel. We
assume periodical boundary conditions and perform the spatial
integrations in Fourier space, which allows us to also account for
radiation with mean free paths larger than the box size. In the
calculation, large scale structure is tracked from $z \sim 60$, just
after the first stars are expected to form \citep{Naoz:2006,
  Fialkov:2012}, to $z \sim 7$ when neutral gas is completely
reionized; specifically, we assume a total optical depth to
reionization of $\tau = 0.075$, our ``late'' reionization case from
\citet{Fialkov:2014b} which is motivated by various observations
\citep{WMAP,Ade:2013,z6}.  Simultaneously we evolve three-dimensional
cubes of inhomogeneous X-ray, \Lya, and LW radiative
backgrounds as well as the ionized fraction of gas (accounting for the
partial ionization by X-rays in regions not yet reionized by stellar
ultra-violet photons). We average all our results over six data cubes
with different randomly generated realizations of the initial
conditions for density and relative velocity fields. This averaging
brings us closer to expected real measurements since the field of view
in realistic observations is significantly larger than even our fairly
large box size.

Following \citet{Fialkov:2014b}, we further explore the impact of the
realistic hard X-ray power spectrum on the 21-cm signal, expanding the
redshift range to $7<z<40$. We compare the 21-cm signal in the case of
gas heated by sources with the soft (previously conventional)
power-law X-ray spectrum to the case of gas heated by a realistic
population of X-ray binaries. We consider three possible histories of
star-forming halo populations: (1) ``massive halos'', which assumes
that stars form in halos with circular velocities above $V_c = 35$ km
s$^{-1}$ ($M \gtrsim 3\times 10^8 M_\odot$ at $z = 20$), an example of
the case where lower-mass halos are inefficient at star formation,
e.g., due to internal feedback from supernovae or a mini-quasar; (2)
``atomic cooling halos'', in which the minimum mass for cooling and
star formation is set by the need for atomic hydrogen to be
radiatively cooled, i.e., $V_c = 16.5$ km s$^{-1}$ and above ($M
\gtrsim 3\times 10^7 M_\odot$ at $z = 20$); (3) ``molecular cooling
halos'' - halos with masses down to the minimum halo mass in which
stars form through molecular cooling, i.e., $V_c\gtrsim 4$ km s$^{-1}$
($M \gtrsim 7\times 10^5 M_\odot$ at $z = 20$ if there were no
feedback). We add this third case since we are considering a broad
redshift range including very early times in which the smallest halos
may have dominated star formation.

In the molecular cooling case, the population of halos is suppressed
by the negative feedback of the Lyman-Werner (LW) photons; motivated
by the simulation results of \citep{Visbal:2014}, we assume the
``strong'' LW feedback case from \citet{Fialkov:2013} and
\citet{Fialkov:2014}.  In addition, these light halos suffer
significantly from the effect of the relative velocity $\vbc$, which
includes a suppression of halo abundance and gas fraction, and an
increase in the minimal cooling mass (e.g., \citet{Tseliakhovich:2010,
  Dalal:2010, Tseliakhovich:2011, Greif:2011, Stacy:2011,
  Fialkov:2012}).  We use our semi-numerical methods
\citep{Visbal:2012, Fialkov:2012, Fialkov:2013, Fialkov:2014} to
include these effects as we evolve in the same framework the
large-scale distribution of stars and the radiative backgrounds
emitted by these stars. We use the standard set of cosmological
parameters \citep{Ade:2013}, a star formation efficiency of $f_* =
0.05\%$ (with additional log(M) suppression at small masses
\citep{Machacek:2001}) for the case of the molecular cooling halos,
$f_* = 0.05\%$ for the case of the atomic cooling halos and $f_\star =
0.15\%$ for the massive halos. In each case $f_*$ is chosen so that the
escape fraction of ionizing photons would need to be $\sim 20\%$ for a
reionization history with our assumed total optical depth to
reionization of $\tau = 0.075$. LW parameters are adopted from
\citet{Fialkov:2013}, with \Lya   ~parameters from
\citet{Barkana:2005b} including the wing-scattering correction from
\citet{Naoz:2008} and the $\vbc$-dependent suppression of the
filtering mass from \citet{Naoz:2013}.  As noted above, we compare two
spectral distributions of X-ray photons (shown in Figure~1 of
\citet{Fialkov:2014b}): the ``conventional'' soft power-law with a
spectral index $\alpha = 1.5$ (e.g., as in \citet{Furlanetto:2006b})
and the more realistic hard spectrum of HMXBs from
\citet{Fragos:2013}. The two spectral distributions are calibrated to
the same total ratio of X-ray luminosity to star-formation rate (SFR),
$L_X/\textrm{SFR} = 3\times 10^{40}$ erg s$^{-1}$ M$^{-1}_{\odot}$ yr,
summed over photon energies in the range 0.2-100 keV. Since the hard
spectrum is more realistic, we limit the soft spectrum runs to the
case of atomic cooling halos.

In this paper (unlike our previous works) we account in the
calculation of the power spectrum of the 21-cm signal for the redshift
space distortions. The hydrogen column that contributes within the
21-cm line width is inversely proportional to the gradient along the
line of sight of the line-of-sight component of the velocity. Within
linear theory the peculiar velocity in Fourier space is given by ${\bf
  \tilde{v}}(\bf k, z) = \frac{i \bf k}{k^2} \tilde{\delta}$, where
$\tilde{\delta}$ is the Fourier transform of the density perturbation
$\delta$ ($\tilde{\delta}$ is a function of the wavevector $\bf k$ and
redshift). Thus, adding the velocity-gradient fluctuation to the
otherwise isotropic contrast of the 21-cm signal $\delta_{21,{\rm
    iso}} \equiv \frac{\delta T_b}{\bar{T}_b}$ (where $\bar{T}_b$ is
the cosmic mean brightness temperature), adds an anisotropic term
$\mu^2 \tilde{\delta}$ \citep{Barkana:2005a}, where $\mu = \cos\theta$
and $\theta$ is the angle of the wavevector with respect to the line
of sight; we denote by $\delta_{21,{\rm iso}}$ the isotropic part of
the 21-cm fluctuatation signal neglecting the perturbation of the
velocity-gradient. As a result of this modification, the power
spectrum of the 21-cm signal, $P(k)$, acquires two additional
(anisotropic) contributions \citep{Barkana:2005a}
\begin{equation}
P(k) = \Piso+P_X+P_\delta,
\end{equation}
where $\Piso$ is the isotropic power spectrum of $\delta_{21,{\rm
    iso}}$ which contains the information about stellar sources and
radiative backgrounds; $P_X$ is $2 \mu^2$ times the cross-correlation
of $\delta_{21,{\rm iso}}$ and $\delta$; and finally $P_\delta$ is
$\mu^4$ times the power spectrum of $\delta$, which contains
cosmological information about the initial conditions for structure
formation.

In principle, each of these contributions can be measured separately
\citep{Barkana:2005a} using their different angular dependence.
Detection of these separate terms would allow multiple constraints on
astrophysics at high redshifts as well as on fundamental cosmology.
When we plot these terms below, we include in $P_X$ a factor of
$\langle 2 \mu^2 \rangle = 2/3$ and in $P_\delta$ a factor of $\langle
\mu^4 \rangle= 1/5$. Including these angle averages is important since
the overall size determines how hard it will be to measure each term.
We also multiply the dimensionless power spectra by a factor of
$\bar{T}_b^2$ to get dimensional spectra in mK$^2$ units like the
spectra that will actually be measured.

\section{Results}
\label{Sec:results}

\subsection{Heating by hard X-rays}

Following the discussion in \citet{Fialkov:2014b} we note that soft
X-rays, such as in the case of the power-law spectrum, have most of
their energy emitted in photons below $\sim 1$ keV. In general, the
comoving mean free path of an X-ray photon is given by
\citep{Furlanetto:2006}
\begin{equation}
\label{eq:mfp}
\lambda_X \approx 4.9\, \bar x_{HI}^{1/3}\left(\frac{1+z}{15}\right)^{-2}
\left(\frac{E}{300~\textrm{eV}}\right)^3\, {\rm Mpc}\ ,
\end{equation}
where $x_{HI}$ is the neutral fraction.  Thus, soft X-ray photons have
relatively short mean free paths, typically several comoving Mpc. This
means that not much energy is lost to cosmic expansion and the gas is
heated efficiently. On the other hand, the mean free path of hard
photons, whose spectrum peaks at $\sim 3$ keV, exceeds the horizon. In
other words, the majority of the hard X-rays travel long distances
before they are absorbed and can transfer their energy to gas, in
which case much of their energy is lost due to redshift effects, and
the remaining energy is absorbed after a long delay. In particular,
most of the photons above a critical energy of $\sim 1$ keV (which is
a roughly redshift independent threshold) have not yet been absorbed
by the start of reionization. Thus, in the case of the hard spectrum,
heating is much less efficient, and the absorbed energy is reduced by
a factor of $\sim 5$ \citep{Fialkov:2014b}.

\begin{figure}
\includegraphics[width=3.4in]{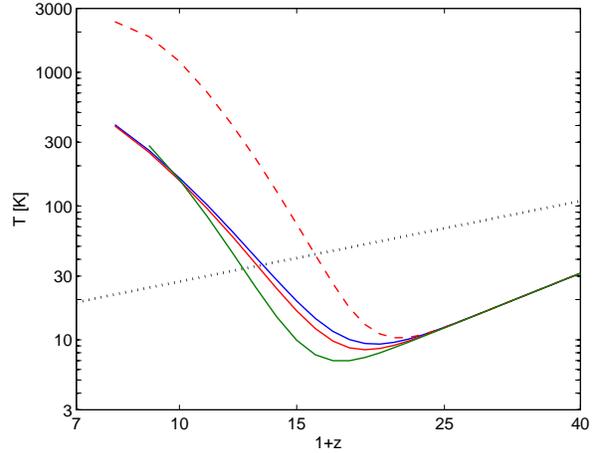}
\caption{The cosmic mean gas temperature. We show the kinetic gas
  temperature (averaged over regions that have not yet been reionized)
  in the cases of molecular cooling halos (blue), atomic cooling halos
  (red) and massive halos (green). We assume the realistic hard X-ray
  spectrum of X-ray binaries (solid curves), except that in the atomic
  cooling case we also compare to a soft power-law spectrum (dashed
  curve). We show the CMB temperature for comparison (dotted curve).
  We note that the molecular cooling case includes the effect of the
  streaming velocity and strong LW feedback, and that the massive halo
  case has a different star-formation efficiency than the other cases
  in order to match the optical depth from reionization.}
\label{fig:1}
\end{figure}

The most direct consequence of heating by the hard X-rays is that
after the period of adiabatic cooling which ends when the first stars
and their remnants heat the gas, cosmic gas warms up slower and more
homogeneously on average around the Universe. The gas reaches the
temperature of the CMB, a milestone termed as the ``heating
transition'', later than previously expected. Fig. \ref{fig:1} shows
thermal histories for the four cases that we consider. Specifically,
for the case of atomic cooling, for which we consider both the hard
and the soft X-ray spectra, the heating transition occurs at redshift
$z = 12.1$ instead of $z = 15.0$ in the case of the soft spectrum. Due
to this shift of $\Delta z \sim 3$, the heating transition happens
when the gas is partially reionized ($z = 12.1$ corresponds to a mean
ionized fraction of $x_i = 0.14$), while previous expectations were
for a clearer separation between the heating of the universe above the
temperature of the CMB ($x_i=0.036$ at $z = 15$) and the later
reionization.

The variation in the model predictions due to different star formation
scenarios (i.e., including light halos or not) is only $\Delta z \sim
0.8$, with the redshift of the heating transition being $ z = 12.3$
and $z = 11.5$ for molecular cooling and massive halos, respectively.
The differences come from the different halo abundances: the number of
massive halos at $z \gtrsim 20$ is small, so heating is initially slow
in this case, but it later catches up due to the higher assumed star
formation efficiency. In the case of the molecular cooling halos,
there is a lot of early structure formation on the corresponding
scales, so all the radiative backgrounds ramp up early, but the
subsequent time evolution is slower. Thus, the timing of the heating
transition is most sensitive to the spectrum of the X-rays (and to
their uncertain normalization, as we explored in
\citet{Fialkov:2014b}).
 
\subsection{Global 21-cm signal}

\label{Sec:global}

Having cool gas during the first half of reioization has immediate
implications for the properties of the observable 21-cm signal emitted
at $z = 7-15$, as discussed in detail in \citet{Fialkov:2014b}.
However, the slower rise of the gas temperature also affects the
predictions for the global 21-cm signal at higher redshifts. In
particular, the absorption trough, which is the most prominent feature
of the global 21-cm spectrum and whose two steep sides will probably
be the easiest to measure, becomes deeper since the gas has more time
to cool adiabatically after it thermally decouples from the CMB; the
first sources then turn on the 21-cm signal with \Lya ~emission before
they start to warm up the gas.

\begin{figure}
\includegraphics[width=3.4in]{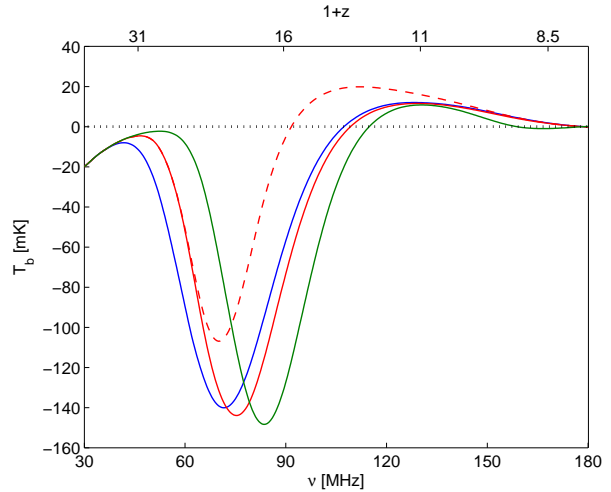}
\caption{The global 21-cm signal. We show the mean 21-cm brightness
  temperature $\bar{T}_b$ relative to the CMB for the cases of
  molecular cooling (blue), atomic cooling (red) and massive halos
  (green). For atomic cooling we also compare to the soft power-law
  spectrum (dashed curve). Quantitative results from this Figure are
  summarized in Table~\ref{Tab:2}.}
\label{fig:2}
\end{figure}

Our results for the global 21-cm signal are shown in
Figure~\ref{fig:2} and summarized in Table~\ref{Tab:2}. As in the case
of the mean cosmic heating history, the variation among the various
cases of star-forming halos  (in terms of the depth of the
  absorption feature which indicates how much the gas cooled before
  being heated by X-rays) is less significant than the difference
between the two X-ray spectra. Specifically, the depth of the
absorption trough is increased by $35\%$, or by $\Delta T_{b} \sim 37$
mK, when the realistic case is compared to the soft power-law spectrum
for the atomic cooling halos, while the variation due to different
star formation histories is only $\Delta T_b = 8$ mK.  Additionally,
the emission peak during reionization is reduced by $42\%$ in the case
of the hard X-rays. This combination of effects implies that global
21-cm experiments should focus on the \Lya ~trough rather than on the
reionization era. We also note that the redshift of the strongest
absorption varies with the halo mass as it depends on the combined
timing of the rise of \Lya coupling and cosmic heating (see also
\citet{Furlanetto:2006} and \citet{Pritchard:2012}).

\begin{table*}
\begin{center}
\begin{tabular}{ |l  | l |l| l|}
  \hline
  Cooling, SED & Absorption minimum & Heating transition 
( $\bar{T}_b = 0$) & Emission maximum  \\
  \hline
  Atomic, Soft X-rays &  $\delta T_b = -107$ mK, $\nu = 70.0$ MHz &  
$\nu = 91.6$ MHz & $\delta T_b = 20$ mK, $\nu = 112.0$ MHz\\
  Atomic, Hard X-rays &  $\delta T_b = -144$ mK, $\nu = 75.5$ MHz &  
$\nu = 109.3$ MHz  &  $\delta T_b = 11.6$ mK, $\nu = 129.0$ MHz\\
  Massive, Hard X-rays &  $\delta T_b = -148$ mK, $\nu = 83.5$ MHz &  
$\nu = 114.8$ MHz  &  $\delta T_b = 10.8$ mK, $\nu = 130.6$ MHz  \\
  Molecular, Hard X-rays  &  $\delta T_b = -140$ mK, $\nu = 71.5$ MHz &  
$\nu = 107.3$ MHz & $\delta T_b = 12$ mK, $\nu = 128.0$ MHz\\
  \hline
\end{tabular}
\caption{\label{Tab:2} Characteristic features of the global 21-cm signal.}
\end{center}
\end{table*}

We note another issue, the small difference between the time at which
the mean gas temperature equals the CMB temperature, and the time at
which the global mean 21-cm temperature is zero. In linear theory
these two times are identical, but in practice non-linearities make
the global spectrum vanish a bit later than the moment of equal gas
and CMB temperatures. With the old spectrum, this delay is $\Delta z
\sim 0.5$ (see also \citet{Fialkov:2013}), but the reduced temperature
fluctuations in the new spectrum reduce the delay to a much smaller
$\Delta z \sim 0.1$. Thus, in what follows, we usually do not consider
both times but instead show results only at the later one (when
$\bar{T}_b = 0$, which is closer to being directly observable), and
refer to this time as the ``heating transition''.
 
\subsection{Fluctuations in the 21-cm signal}

\begin{figure*}
\includegraphics[width=3.4in]{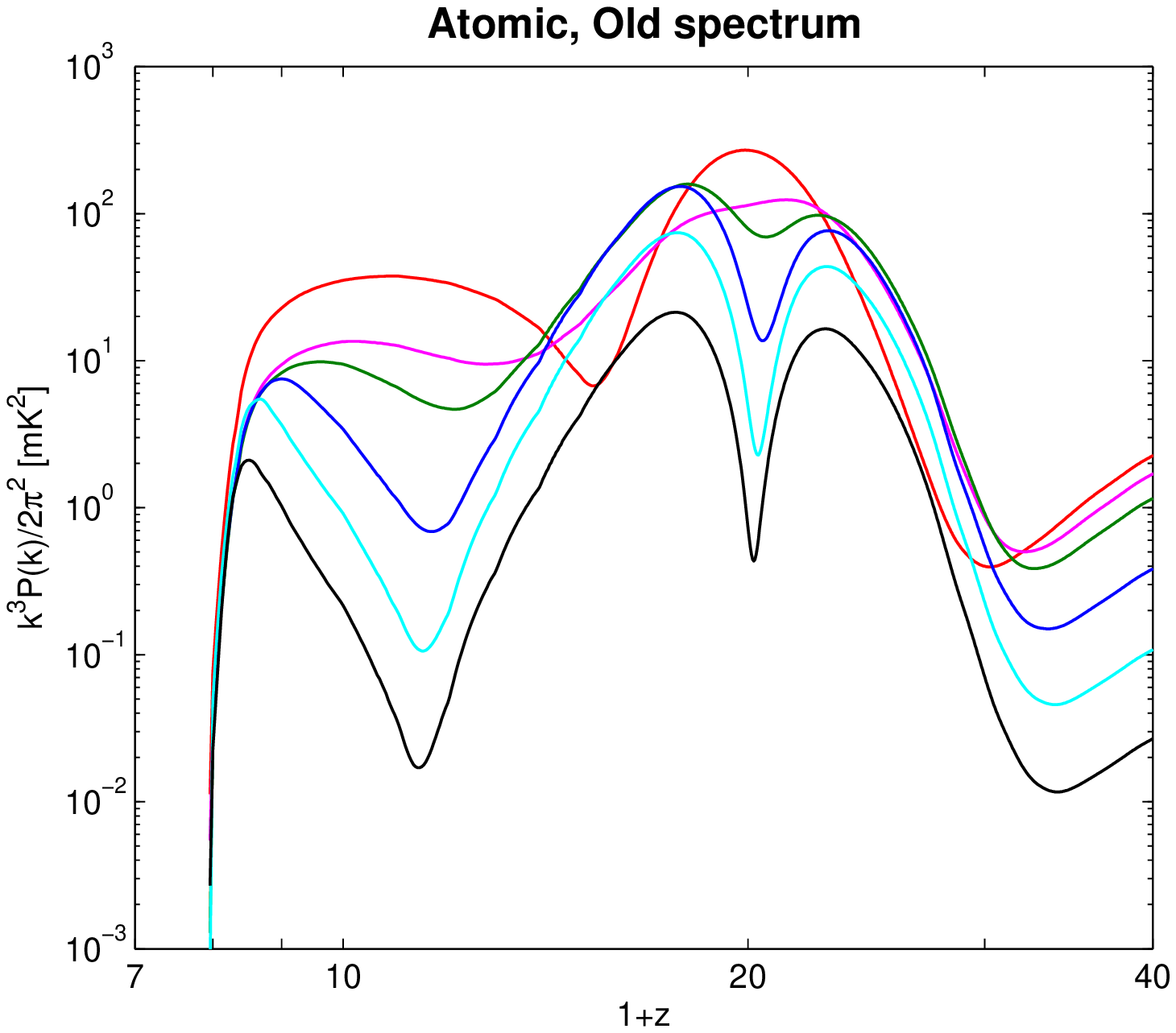}
\includegraphics[width=3.4in]{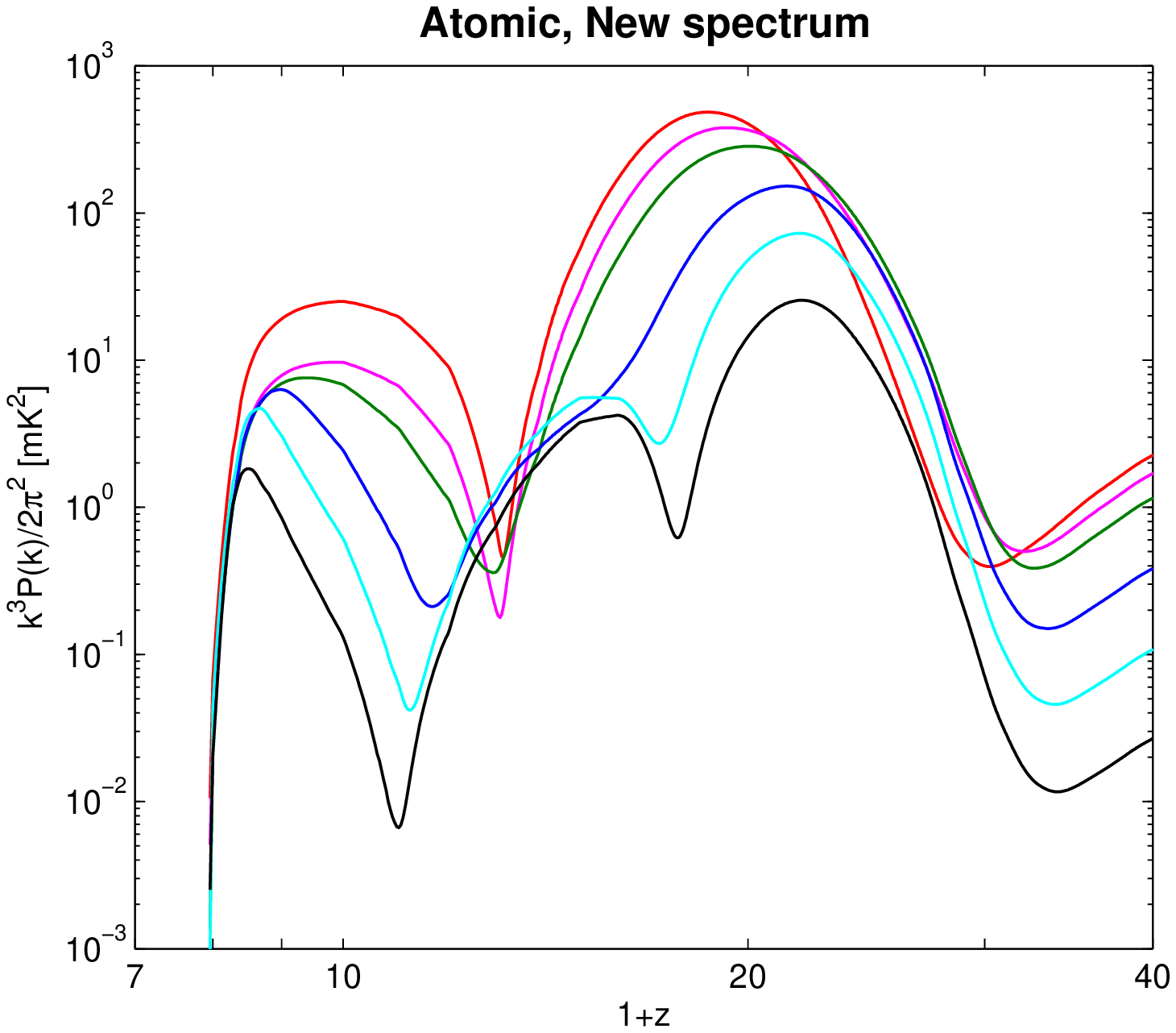}
\includegraphics[width=3.4in]{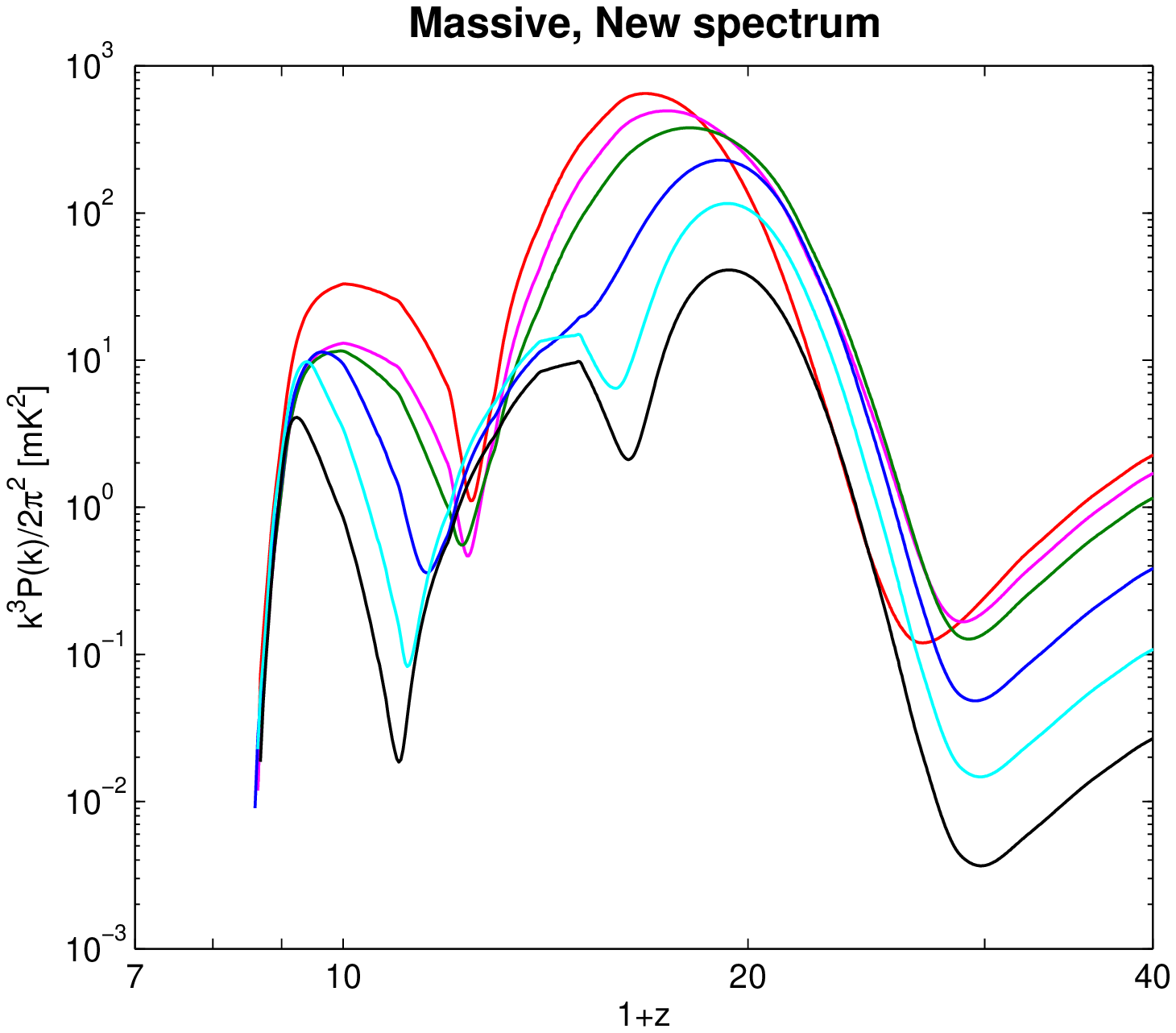}
\includegraphics[width=3.4in]{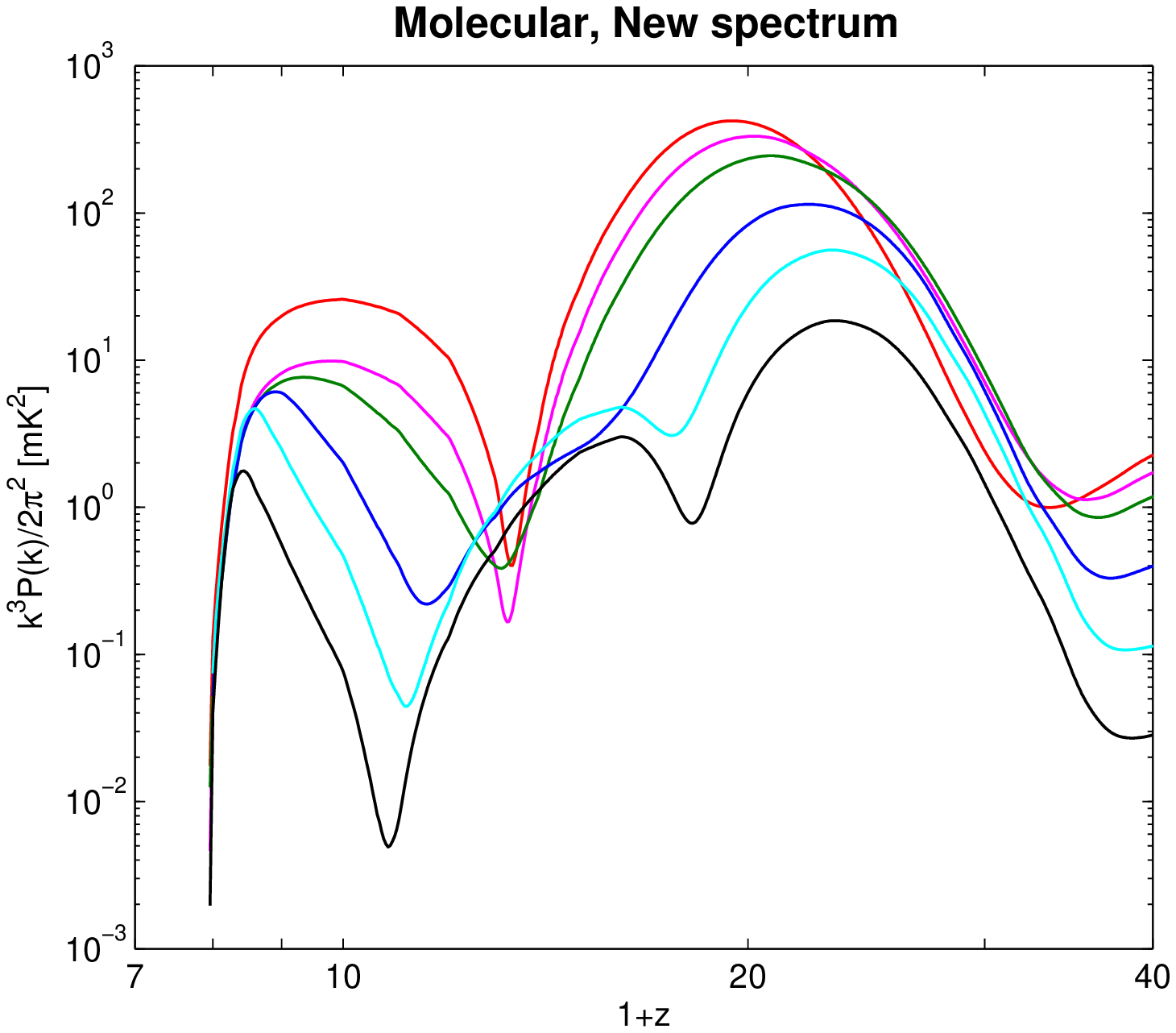}
\caption{Full power spectrum in mK$^2$ units versus redshift at
  different k: 1 Mpc$^{-1}$ (red), 0.5 Mpc$^{-1}$ (magenta), 0.3
  Mpc$^{-1}$ (green), 0.1 Mpc$^{-1}$ (blue), 0.05 Mpc$^{-1}$ (cyan),
  and 0.03 Mpc$^{-1}$ (black). Top left: atomic cooling with old X-ray
  SED. Top right: atomic cooling with new SED. Bottom left: massive
  halos with new SED. Bottom right: molecular halos with new SED.}
\label{fig:3}
\end{figure*}

The power spectrum of the fluctuations in the 21-cm signal contains
much more information than the global signal. Our main focus in this
paper is to explore the whole span of this signal, including broad
ranges in redshift and scale, and the additional terms from the
line-of-sight anisotropy. We begin with Figure~\ref{fig:3}, which
shows the reshift dependence of the power spectrum at various
wavenumbers between 0.03 and 1 Mpc$^{-1}$, for the four parameter
combinations considered here. In addition, we list some of the most
interesting numbers that characterize the power spectra in our four
cases, for the wavenumbers k = 0.05 Mpc$^{-1}$ and k = 0.3 Mpc$^{-1}$
in Table~\ref{Tab:3}. 

\begin{table*}
\begin{center}
\begin{tabular}{ |l  | r |r| r|r| r| r| r|r|r|r|r|r|}
\hline
Cooling, SED, wavenumber  &  Reion. [mK$^2$] & $z$ & Trough [mK$^2$] & 
$z$ & Heating [mK$^2$] & $z$  & \Lya [mK$^2$] & $z$ \\
\hline
Atomic, Soft X-rays, $k$ = 0.3 Mpc$^{-1}$ & 13.5 & 8.6 & 4.7 & 
11.1 & 158.5 & 17.0 & 97.4 & 21.6 \\
Atomic, Soft X-rays, $k$ = 0.05 Mpc$^{-1}$ & 5.5 & 7.7 &0.1 & 
10.5 & 74 & 16.7 & 43.5 & 21.9\\
Atomic, Hard X-rays, $k$ = 0.3 Mpc$^{-1}$  & 8.4 & 7.6 & 0.36 & 
12.0 & - & - & 284 & 19.1 \\
Atomic, Hard X-rays, $k$ = 0.05 Mpc$^{-1}$ &  4.7 & 7.7 & 0.05 & 
10.2 & 5.5 & 14.5 & 72.6 & 20.8\\
Massive, Hard X-rays, $k$ = 0.3 Mpc$^{-1}$  & 11.5 & 9.0 & 0.55 & 
11.2 & - & - & 380 & 17.1 \\
Massive, Hard X-rays, $k$ = 0.05 Mpc$^{-1}$ &  9.8 & 8.4 & 0.1 & 
10.1 & 14.0 & 14.0 & 116 & 18.3\\
Molecular, Hard X-rays, $k$ = 0.3 Mpc$^{-1}$  & 7.6 & 8.3 & 0.4 & 
12.1 & - & - & 245 & 19.8\\
Molecular, Hard X-rays, $k$ = 0.05 Mpc$^{-1}$ & 4.7 & 7.6 & 0.05 & 
10.1 & 4.8 & 15.1 & 56 & 22.1\\
\hline
\end{tabular}
\caption{\label{Tab:3} Detailed features of the 21-cm power spectrum 
shown in Fig.~\ref{fig:3}. The table shows the values of the power
spectrum and the redshift at (from left to right) the reionization
peak, the trough around the heating transition, the heating peak, and the
\Lya peak, for the four considered cases and at two wavenumbers, $k = 0.3$
and $k = 0.05$~Mpc$^{-1}$.}
\end{center}
\end{table*} 

The striking impact of the hard X-rays is that there are no heating
fluctuations at the scales where they are expected to be found when
soft X-rays heat up the gas. In particular, comparing the two top
panels of Fig.~\ref{fig:3}, which show the two cases of atomic cooling
with the soft (left panel) and hard (right panel) X-rays, the peak at
$z \sim 15-20$, which appears in the case of soft X-rays at most of
the scales presented in the figure ($k<0.5$~Mpc$^{-1}$), basically
disappears in the case of hard X-rays. Instead of the three peaks (at
$k<0.5$~Mpc$^{-1}$) which appear due to the inhomogeneous radiative
backgrounds (at $z\sim 22$ from the inhomogeneous \Lya ~background, at
$z\sim 17$ due to X-ray heating and at $z \sim 7.5-8$ due to patchy
reionization) in the case of the soft X-rays (in agreement with the
standard theoretical studies of the 21-cm signal, e.g.,
\citet{Pritchard:2008} and \citet{Mellema:2013}), with a hard SED 
there are only two peaks (at $k>0.05$~Mpc$^{-1}$) contributed by the
fluctuations in the \Lya ~background and the ionization fraction.

This phenomenon is easy to explain by the fact that the heating is
almost uniform in the case of the energetic X-rays emitted by the
HMXBs. There are no heating fluctuations on scales smaller than the
characteristic mean free path of X-ray photons that contributed to the
heating of the gas. On scales larger than the characteristic mean free
path of the photons that can be absorbed before the heating
fluctuations are saturated, the heating peak is restored (although its
height is reduced due to the uniform heating contribution from photons
coming from even larger distances). For instance, in the case of the
hard SED, the characteristic scale is $k \sim
0.05$~Mpc$^{-1}$. Applying the fitting formula of eq.~\ref{eq:mfp}
which relates the mean free path to the photon energy, the scale can
be translated to the characteristic energy of the X-ray photons which
heat the gas, which in this case appears to be $\sim 0.9$~keV. Thus,
detecting the heating fluctuation peak and measuring the wavenumber at
which it disappears should provide unique information about the
characteristic mean free paths, and thus the spectral energy
distribution, of the photons that heat the gas. In case of the soft
spectrum, a heating fluctuation peak should be found on much smaller
scales, but it disappears at $k \sim 0.5$~Mpc$^{-1}$ and above, which
corresponds to a characteristic X-ray energy of $\sim 0.4$~keV (note
that the soft spectrum implemented here has a cutoff at 0.2 keV). Note
that some 21-cm signatures of heating sources with different X-ray
spectra have also been discussed by \citet{Mesinger:2014}.

In addition, in the cases with the hard X-rays, the trough seen in the
power spectra at $z\sim 10-12$ is deeper than in the case of the soft
X-rays by a scale-dependent factor. For instance, Table~\ref{Tab:3}
shows that at k = 0.3 Mpc$^{-1}$ the spectrum goes as low as
0.36~mK$^2$ instead of 4.7~mK$^2$, which means (after applying a
square root) weaker fluctuations by a factor of four. This feature was
extensively discussed in \citet{Fialkov:2014b}.

The same characteristic behaviors listed above are seen in the cases
of the massive halos and the molecular cooling halos, shown in the
bottom row of Fig.~\ref{fig:3}. The only major difference between the
various star formation scenarios here, as in the case of the global
spectrum, is the earlier, but slower, evolution in the case of
molecular cooling halos, resulting also in a slight shift of the
\Lya ~peak toward higher redshifts, and (on the other hand) the 
later but more intense rise of the signal in the case of the massive
halos.
 
\begin{figure*}
\includegraphics[width=3.4in]{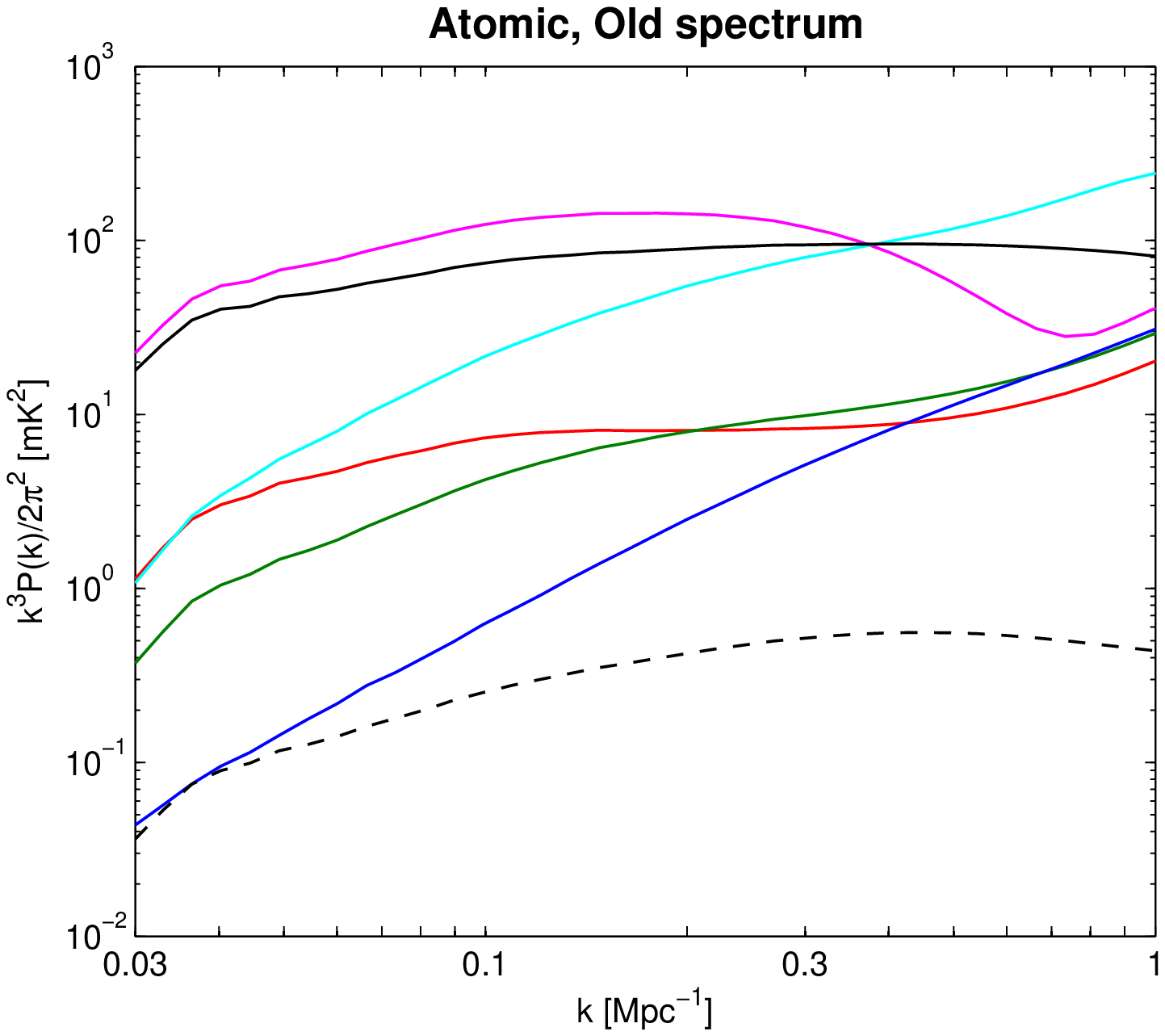}
\includegraphics[width=3.4in]{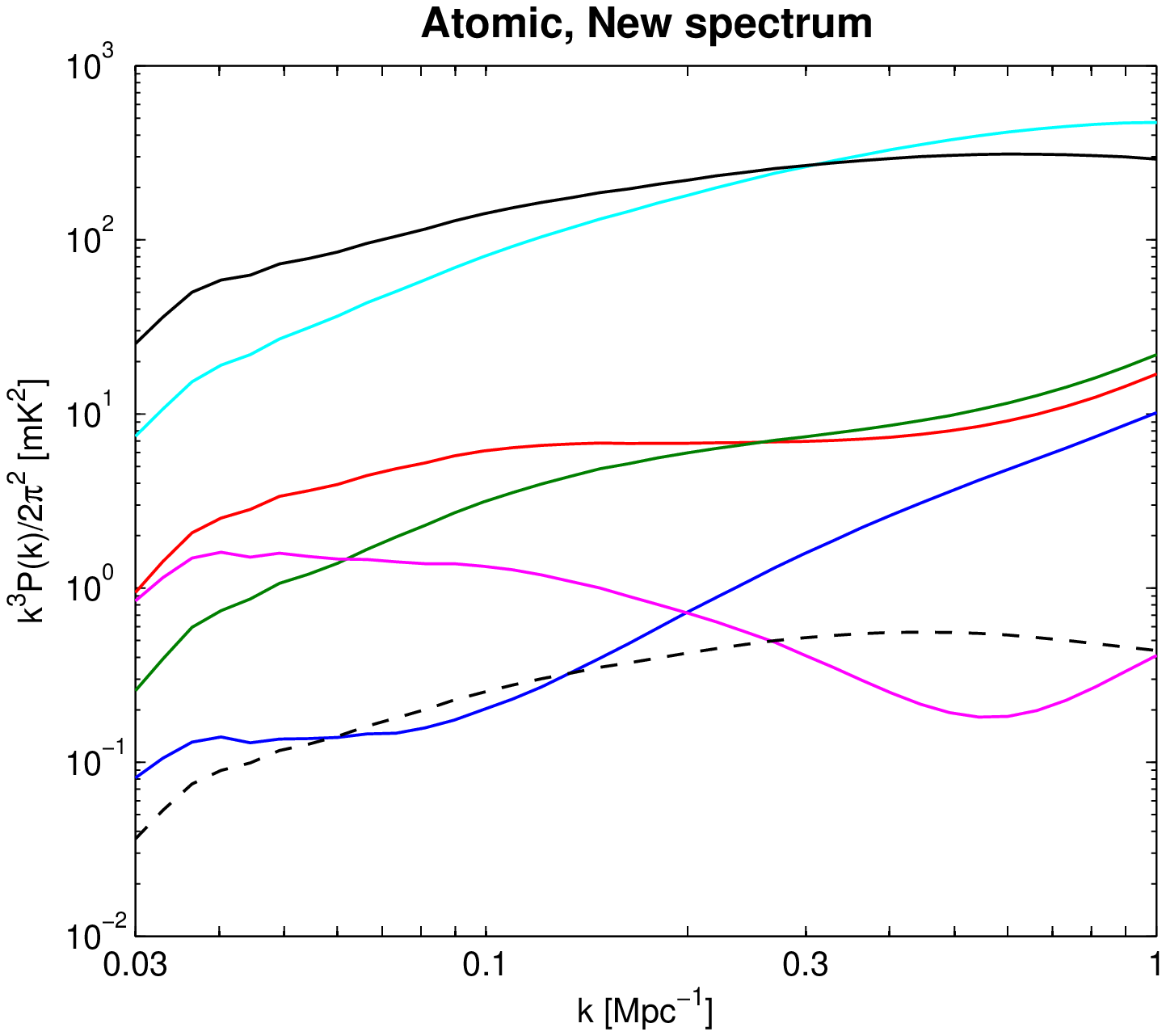}
\includegraphics[width=3.4in]{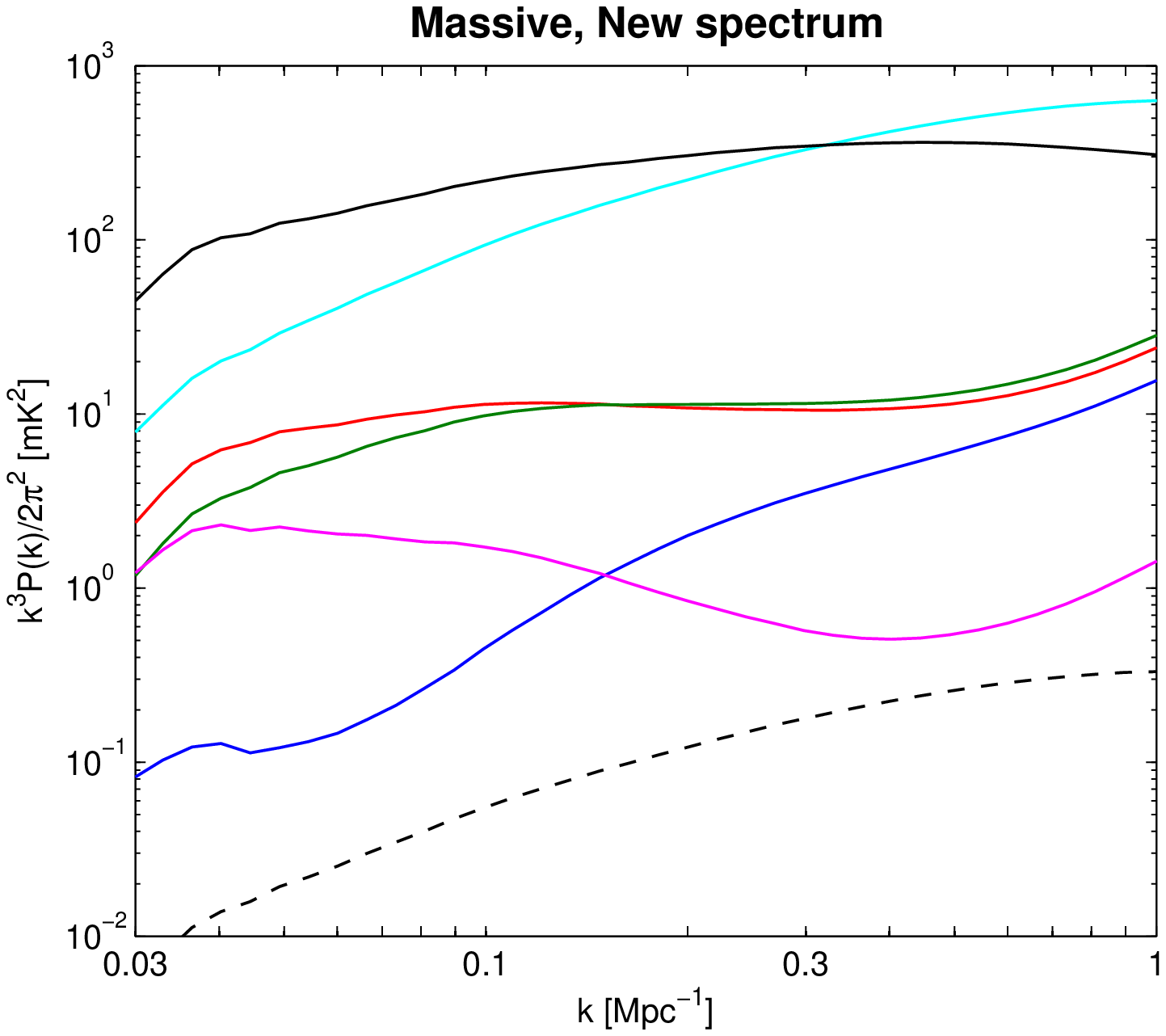}
\includegraphics[width=3.4in]{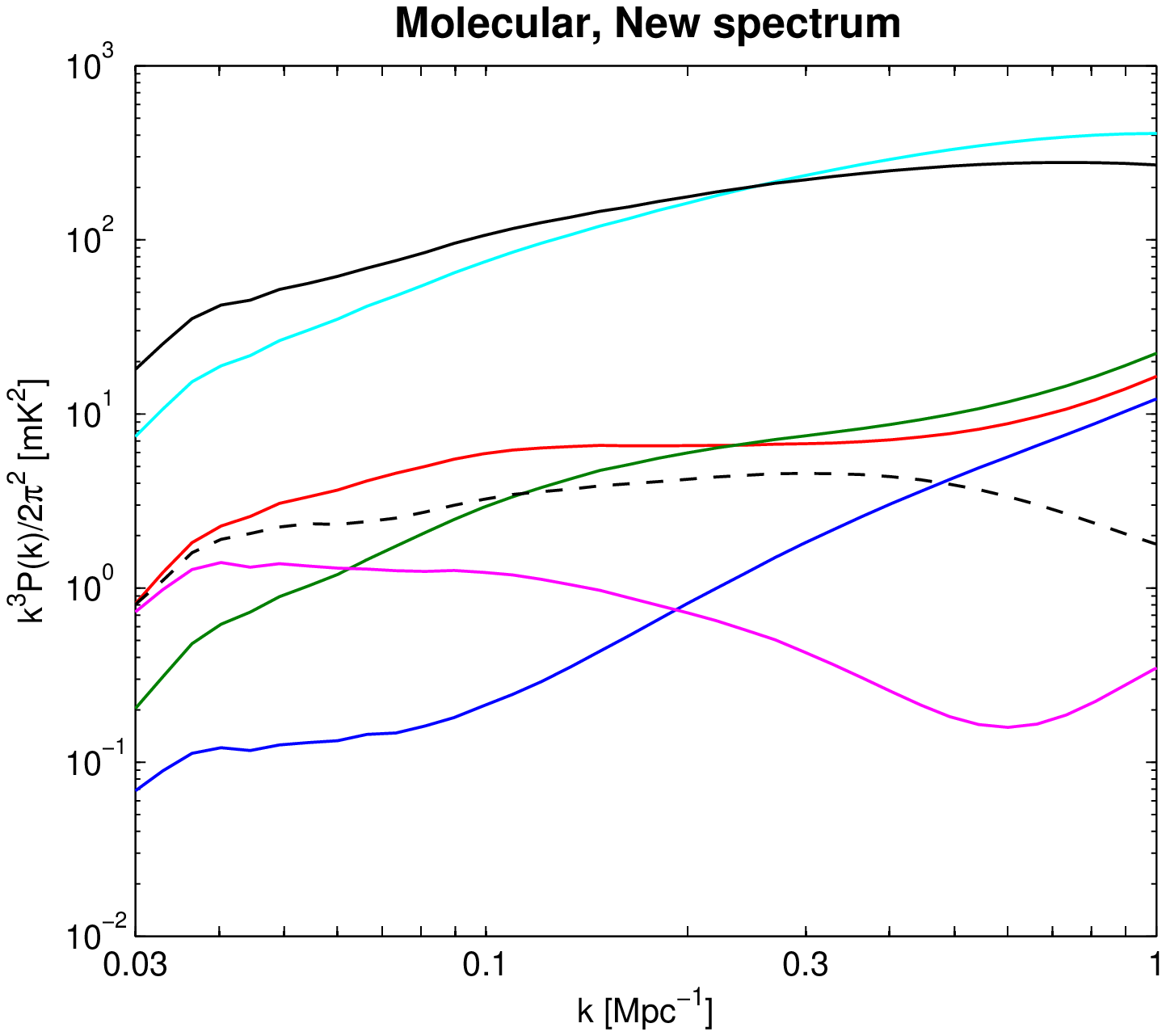}
\caption{Power spectrum in mK$^2$ units versus wavenumber $k$ in 
Mpc$^{-1}$, at several important redshifts (in order from low to
high): that of the reionization peak at $k = 0.1$ Mpc$^{-1}$ (red
solid line), the midpoint of reionization (i.e., $x_i = 0.5$) (green
solid line), $x_i = 0.25$ (blue solid line), the peak of the heating
fluctuations (in the case of the old SED) or of the heating transition
(in the case of the new SED) (magenta solid line), the minimum of the
global $\bar{T}_b$ (cyan solid line), peak of fluctuations from
Ly$\alpha$ (black solid line), and $z = 30$ (black dashed line). Top
left: atomic cooling with old SED, at $z = 8$ (corresponds to $x_i =
0.66$, red), 8.7 (green), 10.7 (blue), 16.8 (magenta), 19 (cyan), 22
(black solid), 30 (black dashed). Top right: atomic cooling with new
SED, at $z = 8$ ($x_i = 0.66$, red), 8.7 (green), 10.7 (blue), 12.1
(magenta), 18 (cyan), 20.4, 30 (black dashed). Bottom left: massive halos with new SED, at
$z = 8.6$ ($x_i = 0.6$, red), 8.9 (green), 10.3 (blue), 11.3 (magenta),
16 (cyan), 18.1 (black solid), 30 (black dashed).  Bottom right:
molecular cooling halos with new SED, at $z = 7.9$ ($x_i = 0.67$,
red), 8.6 (green), 10.6 (blue), 12.3 (magenta), 19 (cyan), 21.2 (black
solid; close to the Ly$\alpha$ peak at 21.7), 30 (black dashed).}
\label{fig:4}
\end{figure*}

In Fig.~\ref{fig:4} we show a different cut of the parameter space:
the total 21-cm power spectrum versus wavenumber, at each of several
key redshifts:  redshift during reionization at which the power
  spectrum magnitude at k = 0.1 Mpc$^{-1}$ reaches its maximum
  (referred to as the reionization peak), the midpoint of reionization
  (i.e., $x_i = 0.5$), the moment when $x_i = 0.25$, the redshift at
  which heating fluctuations peak (in the case of the old SED) or the
  redshift of the heating transition (in the case of the new
  SED)\footnote{For each SED we chose to show the power spectrum at
    the main observable milestone related to the heating era; thus, we
    show the old SED at the peak of heating fluctuations, while the
    new SED (which produces a fluctuation minimum instead of large
    heating fluctuations) is shown at the redshift of the heating
    transition.}, the minimum of the global brightness temperature,
  the redshift at which fluctuations from Ly$\alpha$ peak, and a high
  redshift (here $z = 30$) at which the 21-cm fluctuations are
  dominated by fluctuations in density, although small \Lya\
  fluctuations are already present.  Comparing the two cases of
atomic cooling again, we see that shape of the power spectrum is most
sensitive to the type of the SED at the early stages of reionization
(here shown for $x_i = 0.25$) and during heating, where the key
feature is a peak of heating fluctuations (in the case of the old SED)
or a minimum around the heating transition (in the case of the new
SED). Comparing the various halo masses (all with the new SED), the
more massive halos tend to produce a higher signal due to the higher
halo bias (i.e., stronger clustering), though other factors are also
critical (such as the faster rise of the cosmic star formation rate
for more massive halos, which affects heating and \Lya ~coupling). In
general, the exact shape of the power spectrum has a complex
dependence on the various parameters.

\begin{figure*}
\includegraphics[width=3.4in]{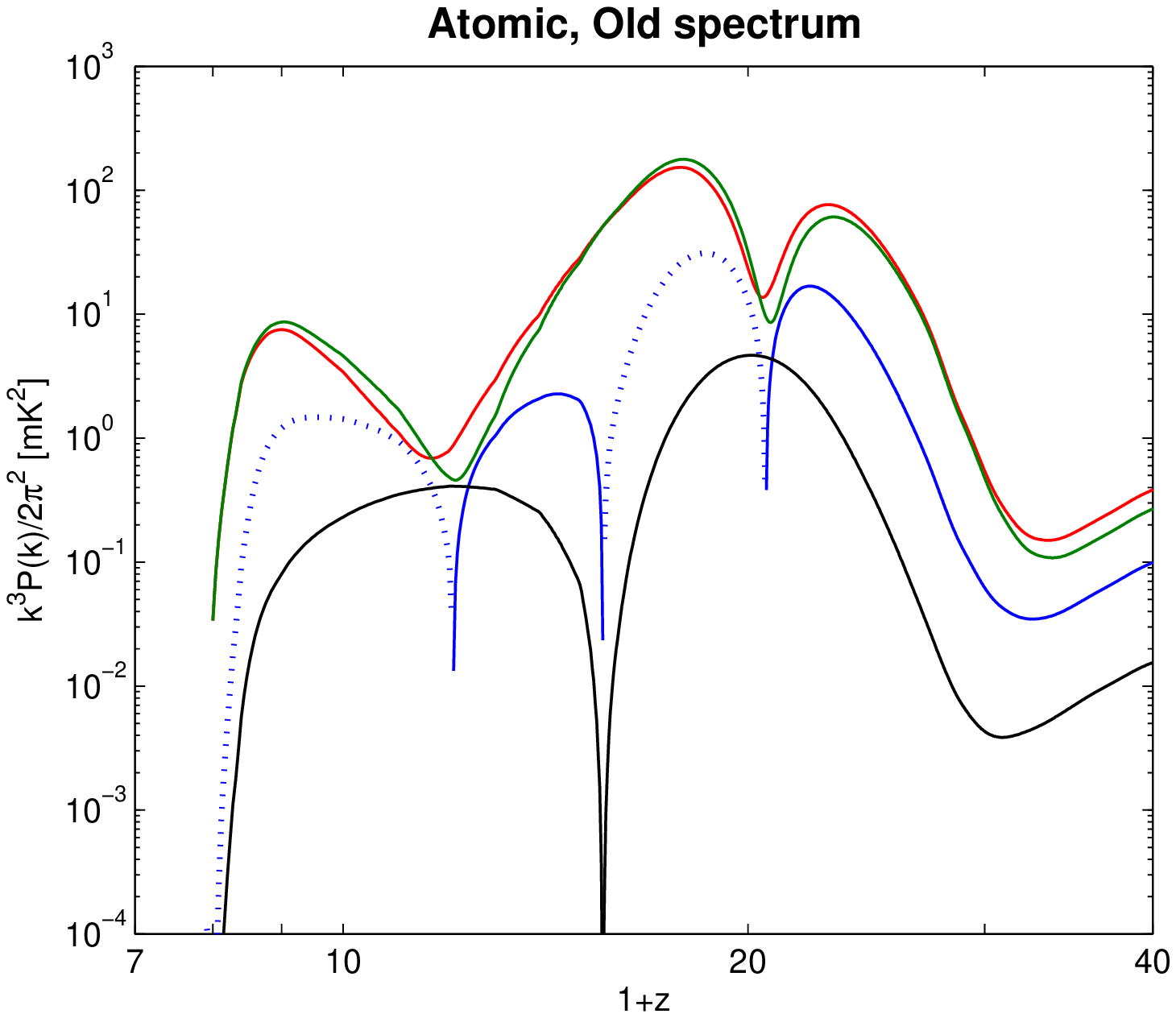}
\includegraphics[width=3.4in]{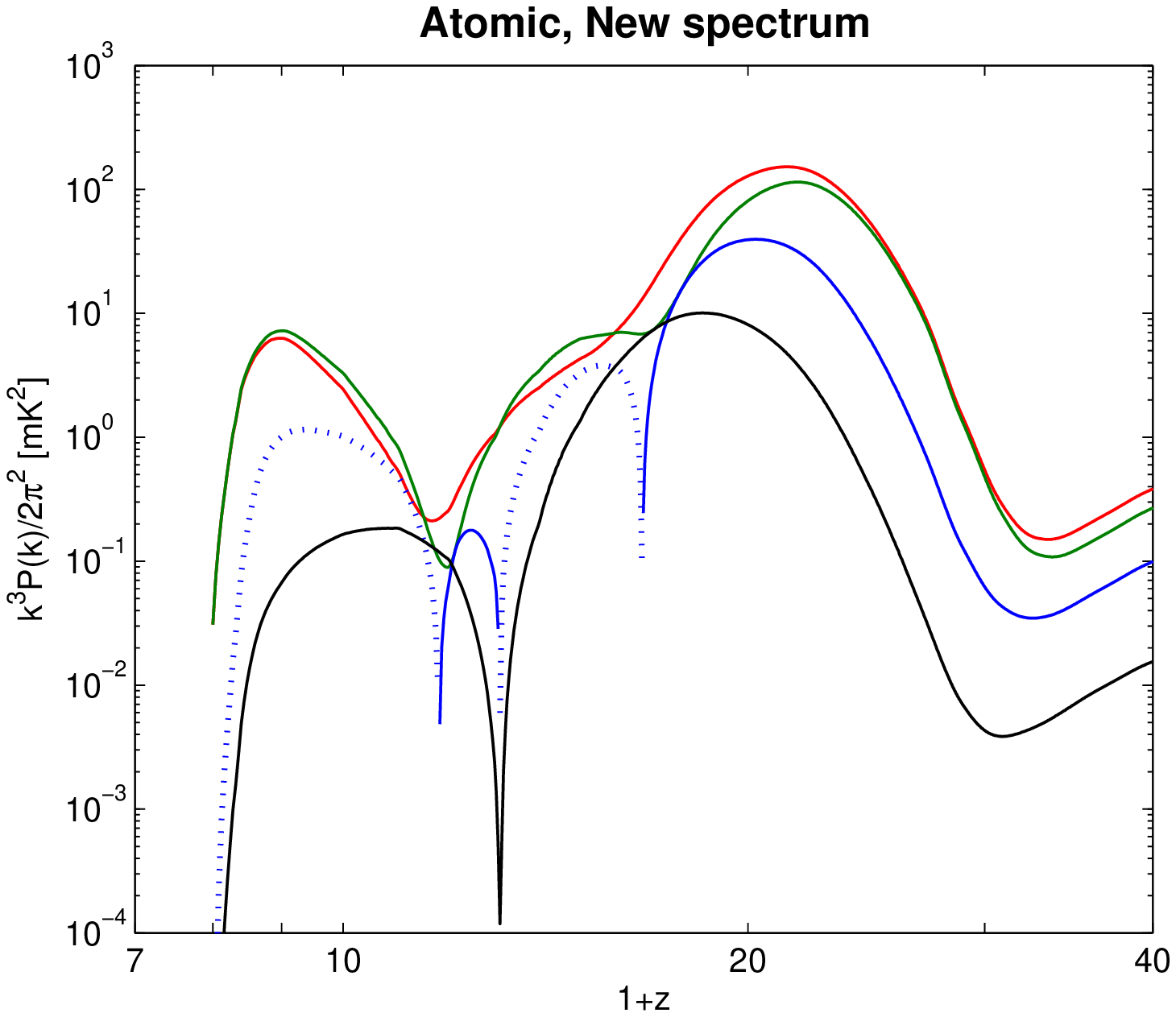}
\includegraphics[width=3.4in]{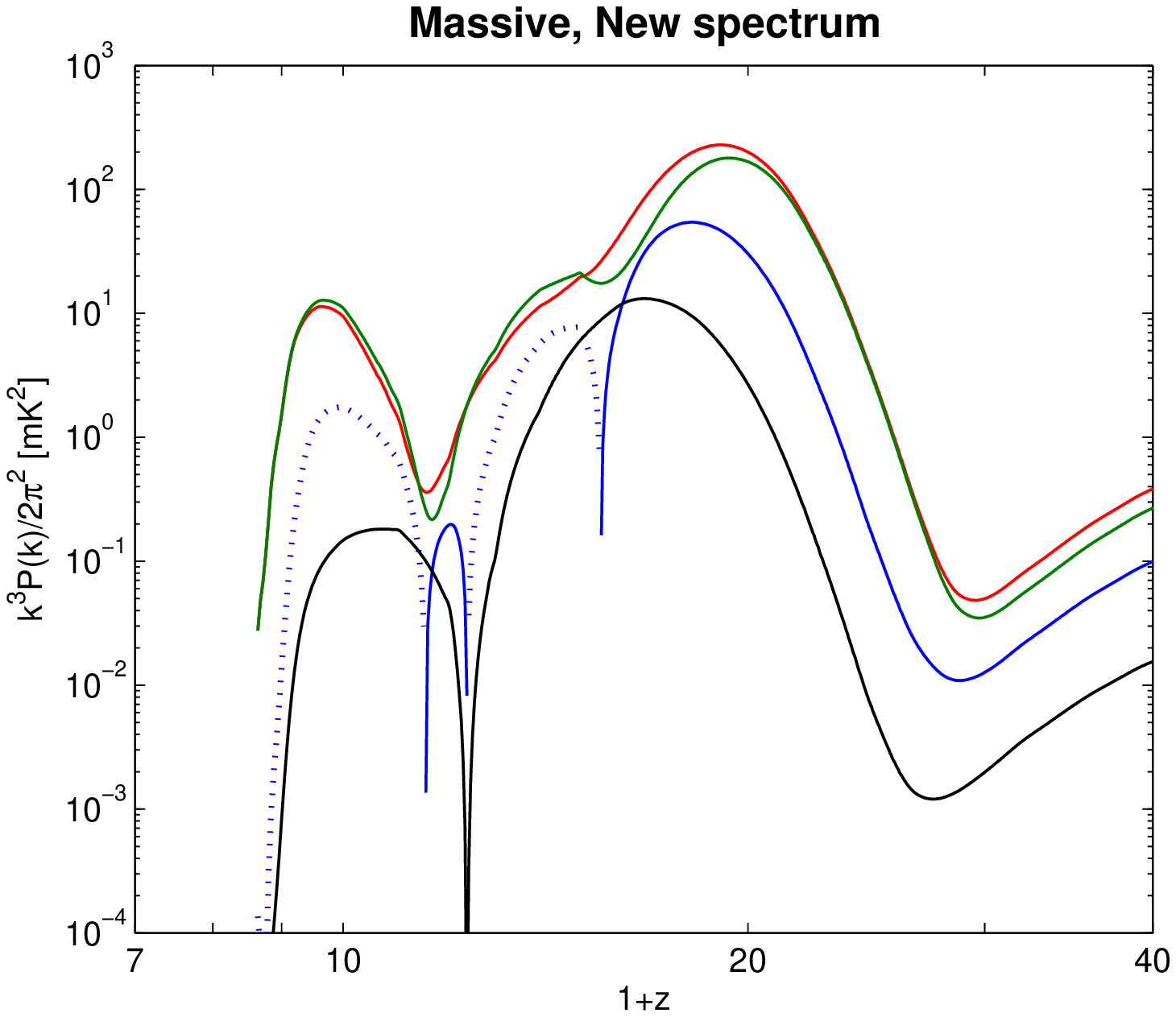} 
\includegraphics[width=3.4in]{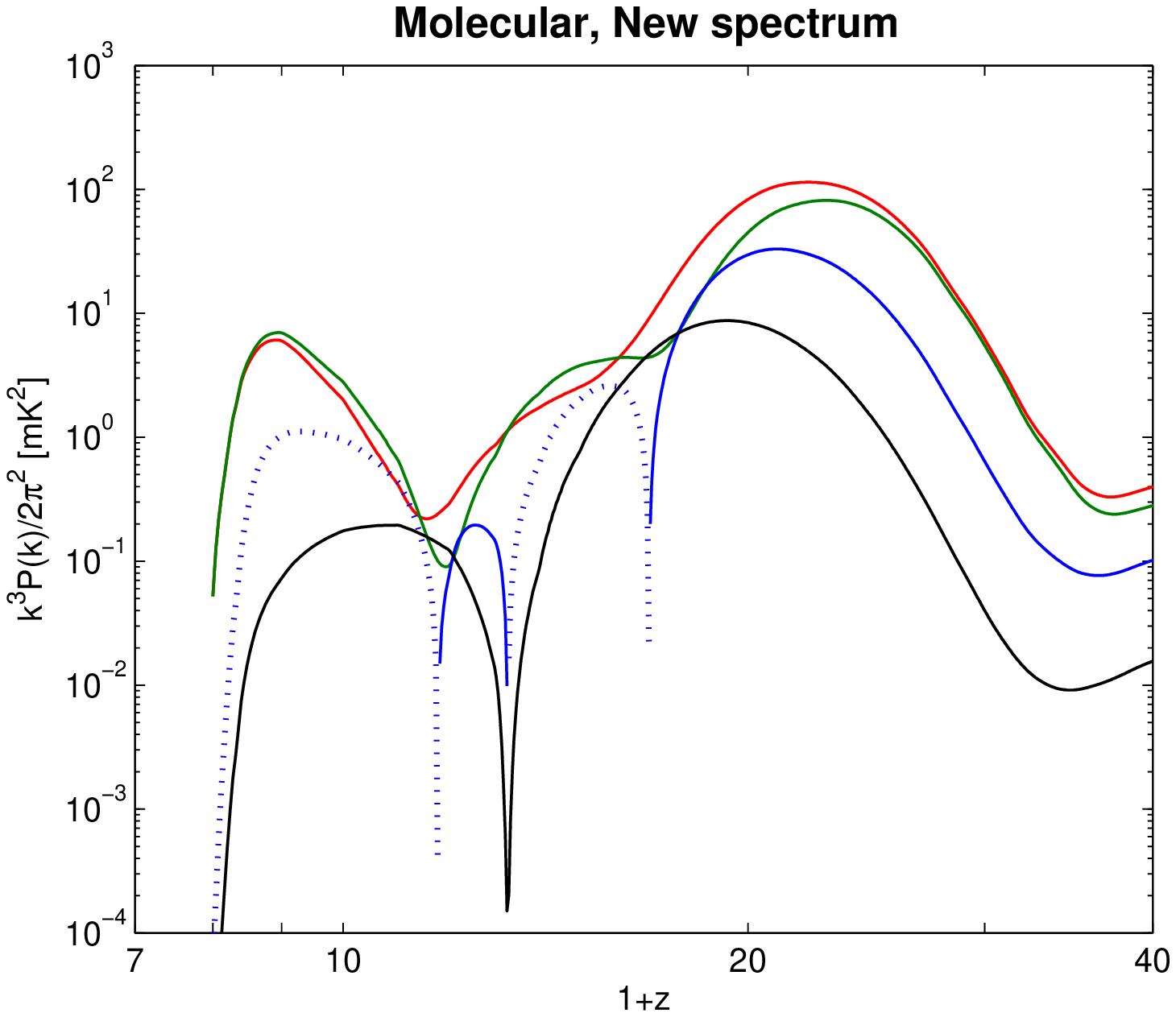}
\caption{Power spectrum in mK$^2$ units at $k = 0.1$~Mpc$^{-1}$ versus
  redshift. We show the full spectrum (red), the isotropic
  contribution $\Piso$ (green), $P_\delta$ (black), and $P_X$ where it
  is positive (solid blue) or $-P_X$ where it is negative (dotted
  blue). Top left: atomic cooling with old SED. Top right: atomic
  cooling with new SED. Bottom left: massive halos with new
  SED. Bottom right: molecular cooling halos with new SED.}
\label{fig:5}
\end{figure*}

As mentioned above, it may be useful to analyze the angular dependence
of the power spectrum in order to learn from the same measurement
about both astrophysics and cosmology \citep{Barkana:2005a,
Pritchard:2008}.  As was shown by \citet{Barkana:2005a} (and as we
mentioned in sec.
\ref{Sec:Methods}), the unique three-dimensional properties of 21 cm
measurements permit a separation of the total power spectrum to
components according to their angular dependence: $P(k) =
\Piso+P_X+P_\delta$. Even if in practice this separation cannot be 
achieved cleanly (we plan to further study this issue), the additional
information from the angular dependence will provide added
astrophysical information and further tests of the predictions of
simulated models. In Fig.~\ref{fig:5} we show the redshift dependence
of each component (also including the total power spectrum) at a
specific wavenumber $k = 0.1$~Mpc$^{-1}$ for the four histories under
consideration. The dominant contribution to the total power comes from
the isotropic component, i.e., astrophysical sources, while the
contribution of the density two-point function $P_\delta$ is generally
smaller by between a factor of a few and 100. On the other hand,
$P_X$, which is the cross-correlation between the isotropic part and
the density fluctuations, is boosted with respect to $P_\delta$ and
can be used to further confirm and constrain the astrophysical
information.

An important qualitative feature is that $P_X$ is positive or negative
at various times, and its sign changes are correlated with changes in
the shape of other components (such as $\Piso$), since they indicate
the dominance of various fluctuation sources. To analyze this, first
note that through the line-of-sight gradient term, a positive density
fluctuation in a given pixel (which yields an infall velocity pattern
that opposes the Hubble expansion and increases the total 21-cm
optical depth) always tends to increase the magnitude of $T_b$ in the
pixel (regardless of its sign). Now, the 21-cm effect of ionization
fluctuations is anti-correlated (since positive density fluctuations
implies more ionization, fewer hydrogen atoms, and thus a lowered
21-cm optical depth). This is why $P_X$ is negative at the lowest
redshifts, where ionization fluctuations dominate the 21-cm signal
(e.g., at $k = 0.1$~Mpc$^{-1}$ as shown in the figure). In this
redshift region, $\Piso$ (as well as the total 21-cm $P(k)$) rises
with time towards the reionization peak, before falling again as
reionization is completed. Now, when the 21-cm fluctuations are
dominated by heating fluctuations, higher density implies stronger
heating, and higher $T_K$ implies higher $T_b$; this is the same as
increasing $|T_b|$ if $T_b > 0$, but it is the opposite if $T_b < 0$.
Thus, $P_X$ during this redshift interval is positive after the
heating transition and negative before it. In particular, the old
spectrum case shows a substantially extended redshift range with
positive $P_X$ since the heating transition occurs much earlier in
this case. Note that while the heating peak (for the soft spectrum) or
minimum (for the hard spectrum) does {\it not}\/ occur right at the
heating transition (except on small scales for the hard spectrum),
there is a predicted transition from negative to positive $P_X$ very
close to the time when $\bar{T}_b=0$. At higher redshifts, when \Lya
~fluctuations dominate the 21-cm fluctuations, $P_X$ is positive since
higher density implies a higher \Lya ~intensity and thus a higher
magnitude of $T_b$. At the highest redshifts, $P_X$ remains positive
as the 21-cm fluctuations become dominated by the direct effect of
density fluctuations along with the associated adiabatic heating.

Looking more broadly at the shape of the curves with redshift, we
again see that in going from the soft to the hard spectrum, the peak
from heating fluctuations disappears from the signals affected by star
formation, i.e., the total power spectrum, the isotropic part and the
cross-term (in $P_X$ there are still two peaks during the heating era,
due to the sign changes explained above, though they are at
substantially lowered fluctuation levels); whereas the impact on
$P_\delta$ is only via the global $\delta T_{b}$ history. We again
note that here $P_\delta$ is given in mK$^2$ units, i.e., it has been
multiplied by the square of $\bar{T}_b$, which for example drives the
power spectrum to zero at the heating transition when the global
spectrum vanishes.

\begin{figure*}
\includegraphics[width=3.4in]{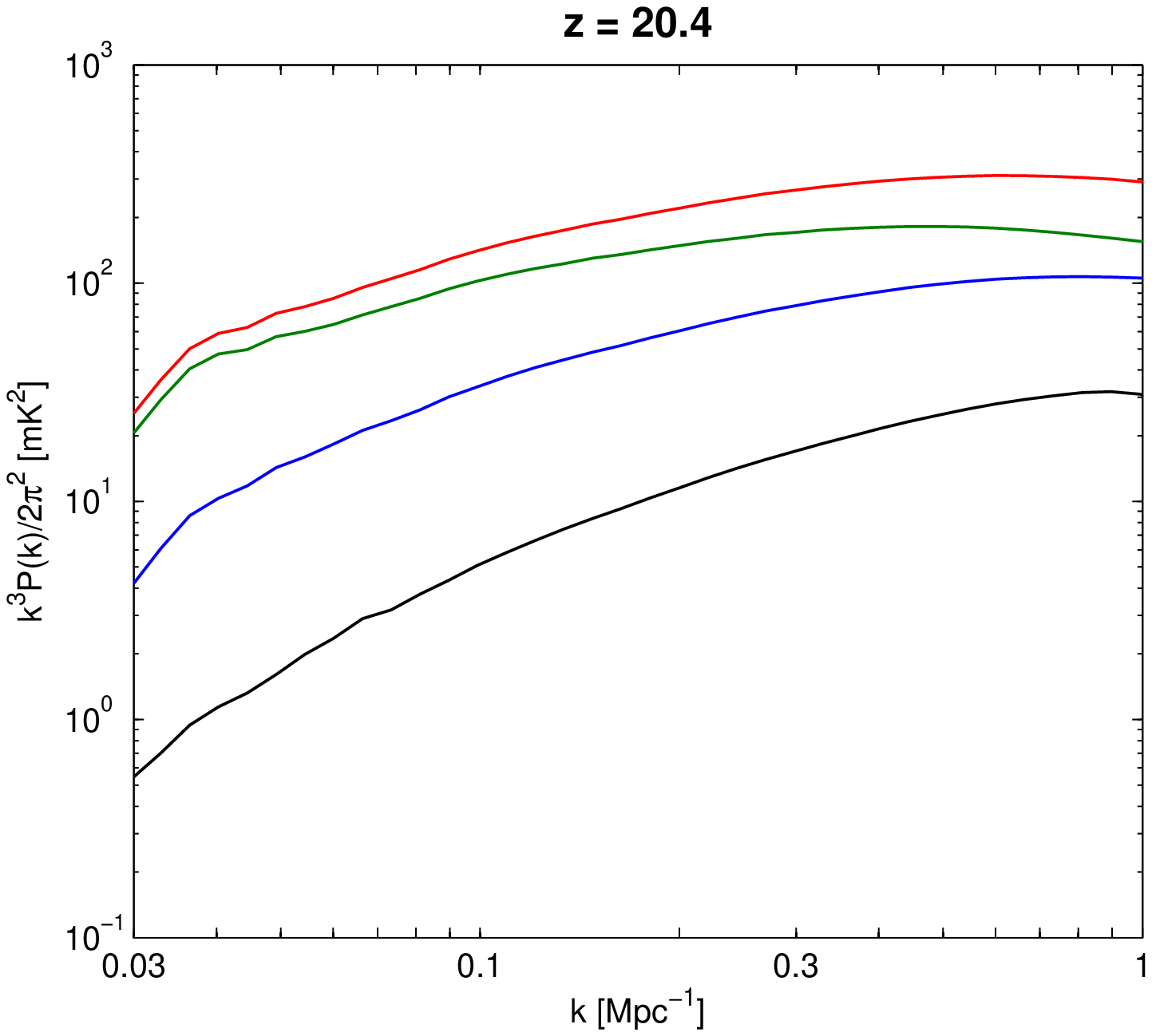}
\includegraphics[width=3.4in]{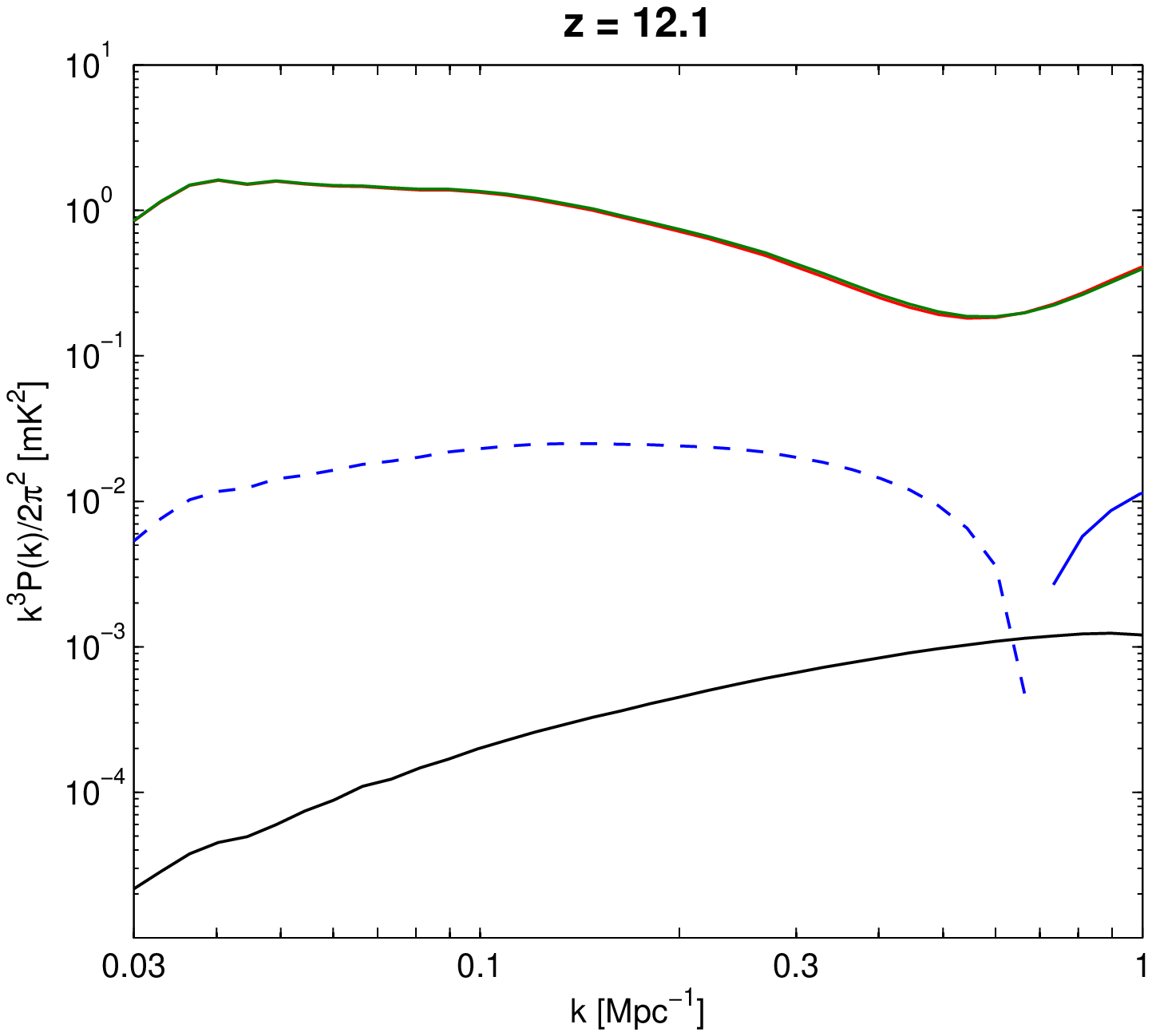}
\includegraphics[width=3.4in]{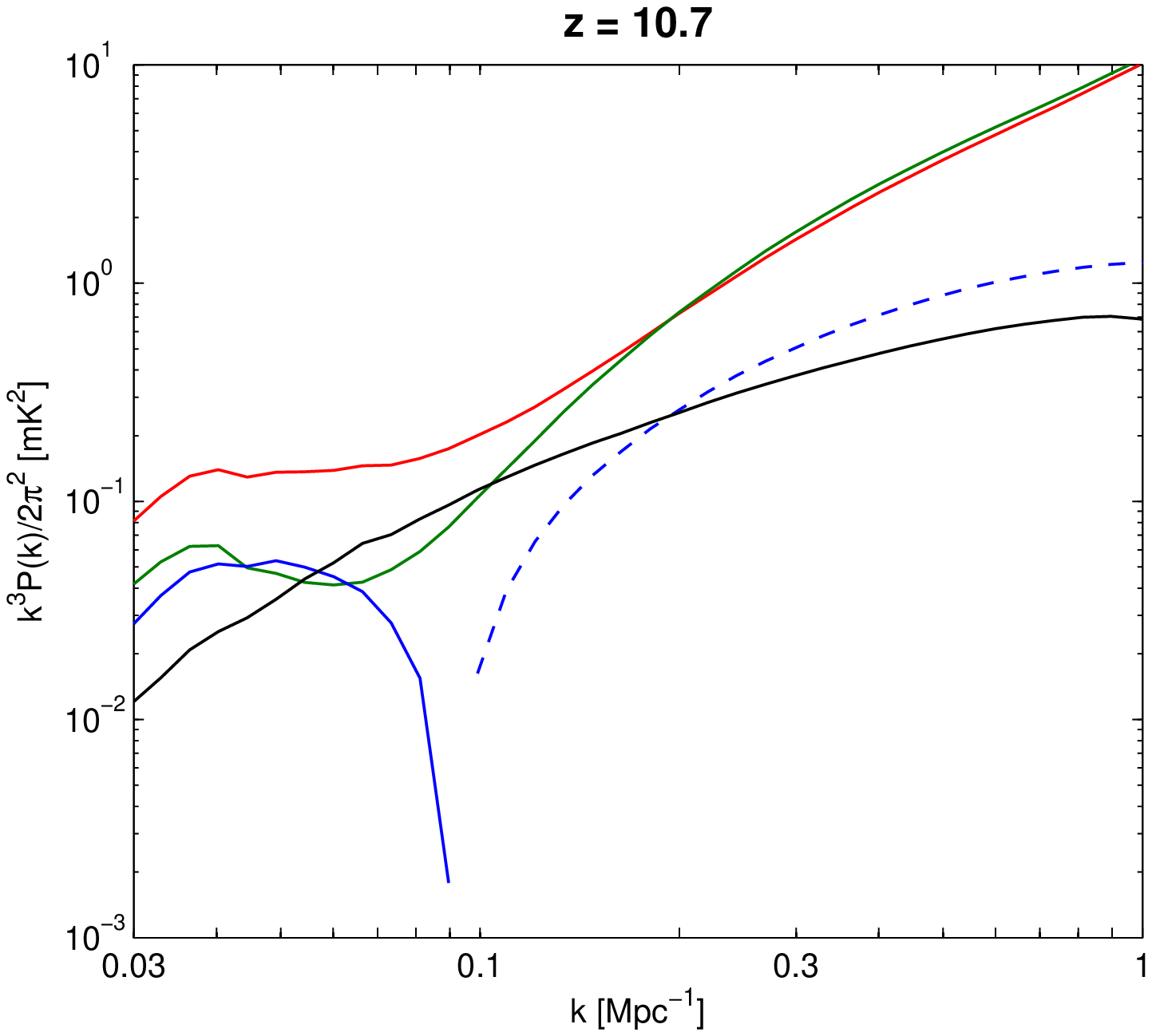}
\includegraphics[width=3.4in]{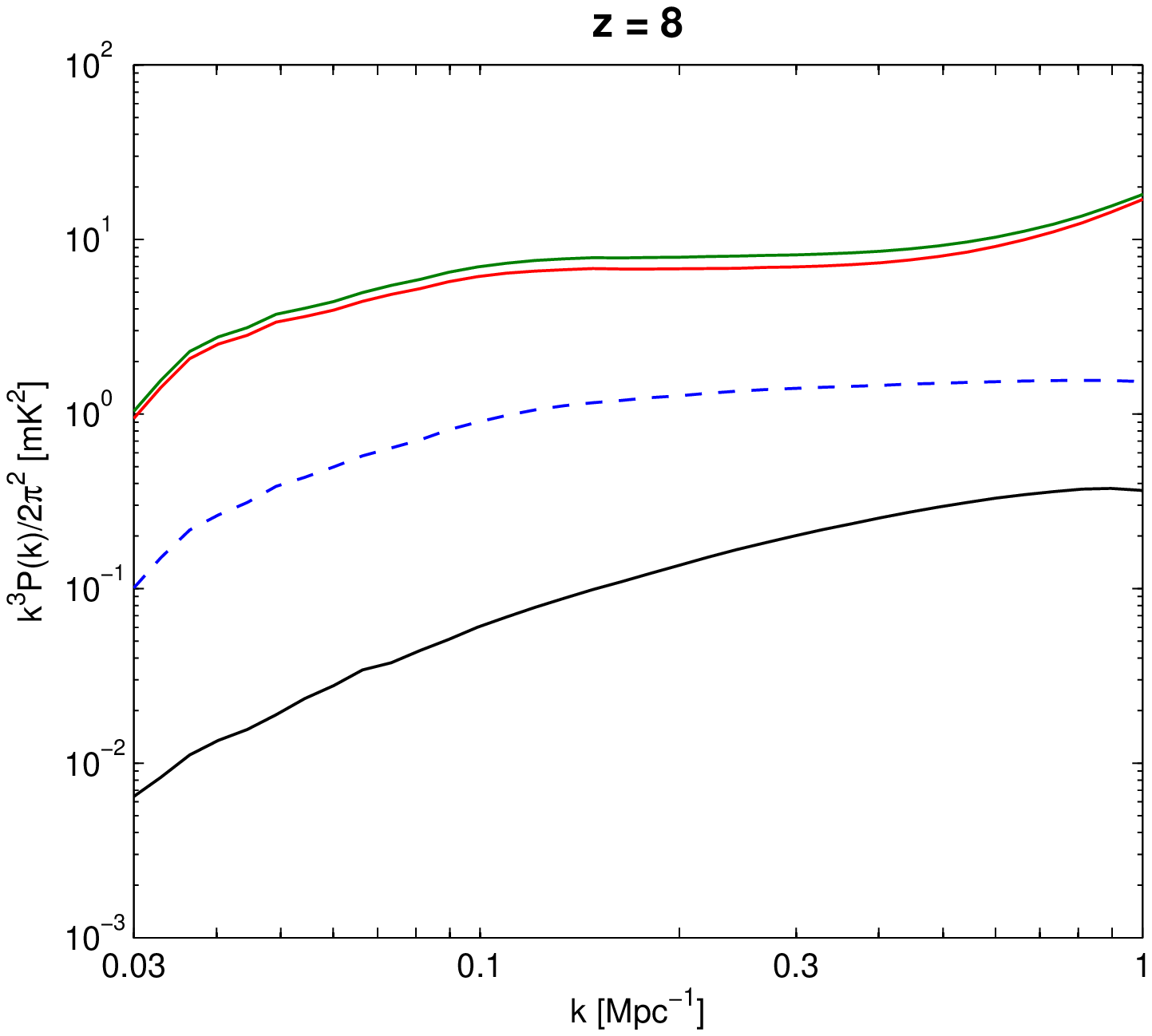}
\caption{Power spectra (in mK$^2$ units) of various components for
  a single parameter case (atomic cooling with the new spectrum)
  versus wavenumber, shown at various redshifts. We show the full
  spectrum (red), the isotropic contribution (green), $P_\delta$
  (black), the positive $P_X$ (solid blue) or minus the negative $P_X$
  (dashed blue). Top left: $z = 20.4$ (the peak of fluctuations
  from Ly$\alpha$). Top right: $z = 12.1$ (heating
  transition). Bottom left: $z = 10.7$ (redshift that corresponds to
  $x_i = 0.25$). Bottom right: $z = 8$ (reionization peak for $k =
  0.1$~Mpc$^{-1}$).  }
\label{fig:6}
\end{figure*}

In order to show the various power spectrum components over a range of
wavenumbers, at least for one case (atomic cooling with hard X-rays),
in Fig. \ref{fig:6} we show the various contributions versus $k$ at
several redshifts which represent different epochs of cosmic history.
This figure shows that the redshifts at which $P_X$ changes sign
depend on $k$, or equivalently, at a given redshift $P_X$ may have a
different sign at different $k$. In particular, at the heating
transition, the fluctuations on most scales are
dominated by temperature fluctuations (for which $P_X>0$ at this
time), but at $z=12$ ionization (with its negative associated $P_X$)
already dominates small scales ($k > 0.6$~Mpc$^{-1}$). As reionization
progresses, ionization fluctuations come to dominate larger and larger
scales, until they dominate the full range of probed scales. At each
redshift (from early to late reionization), the switch in the dominant
source of fluctuations at a particular $k$ is marked both by a switch
in the sign of $P_X$ and in the shape of $\Piso$ (as well as the
total 21-cm power spectrum).

\begin{figure*}
\includegraphics[width=2.6in]{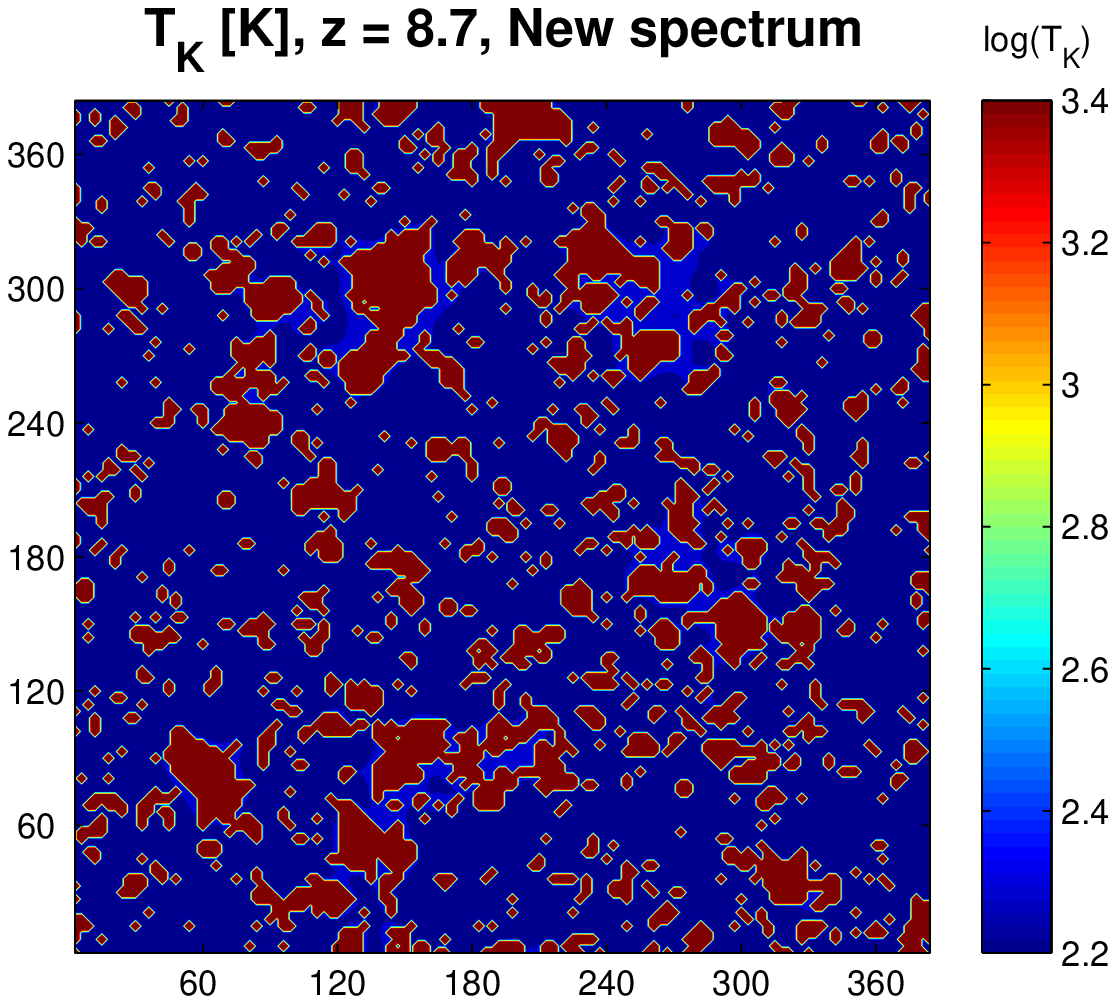}\includegraphics[width=2.6in]{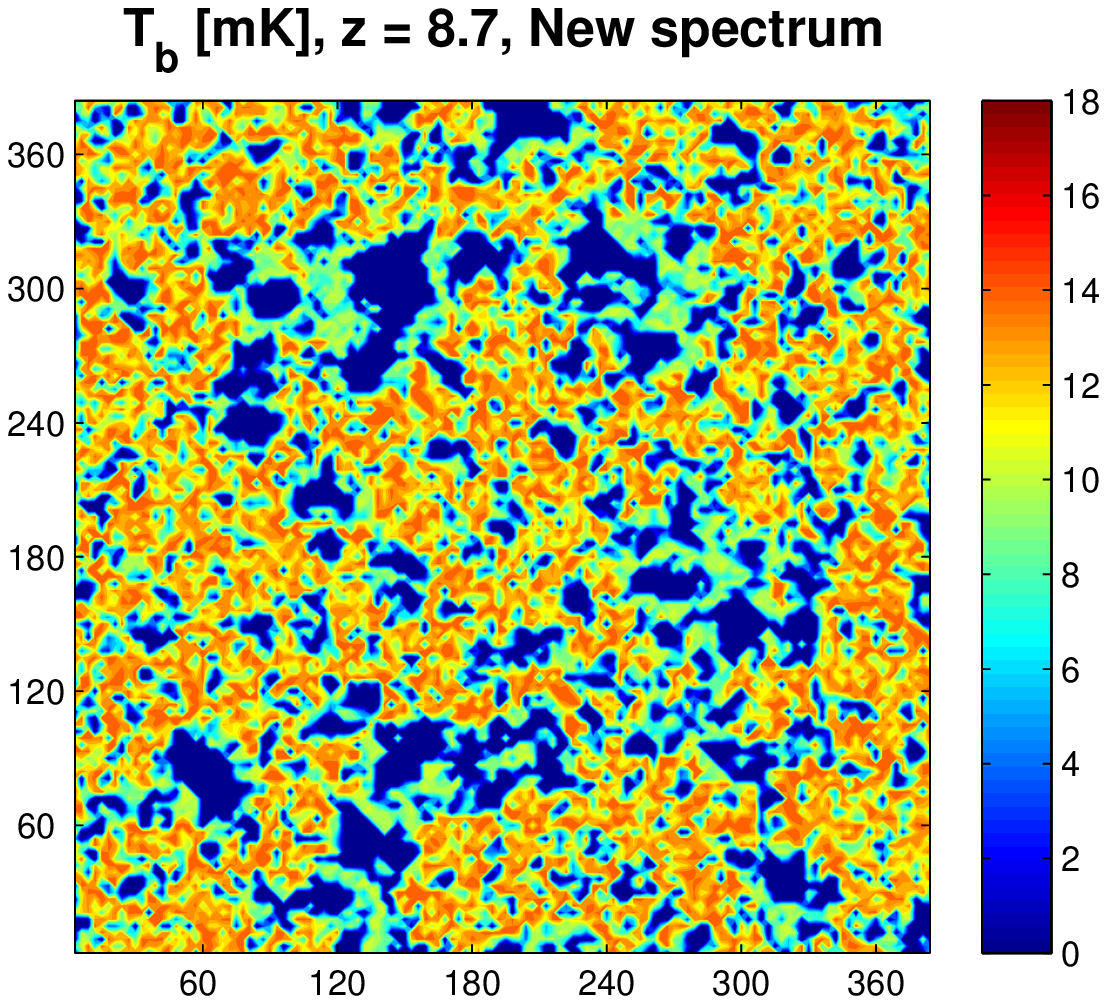}
\includegraphics[width=2.6in]{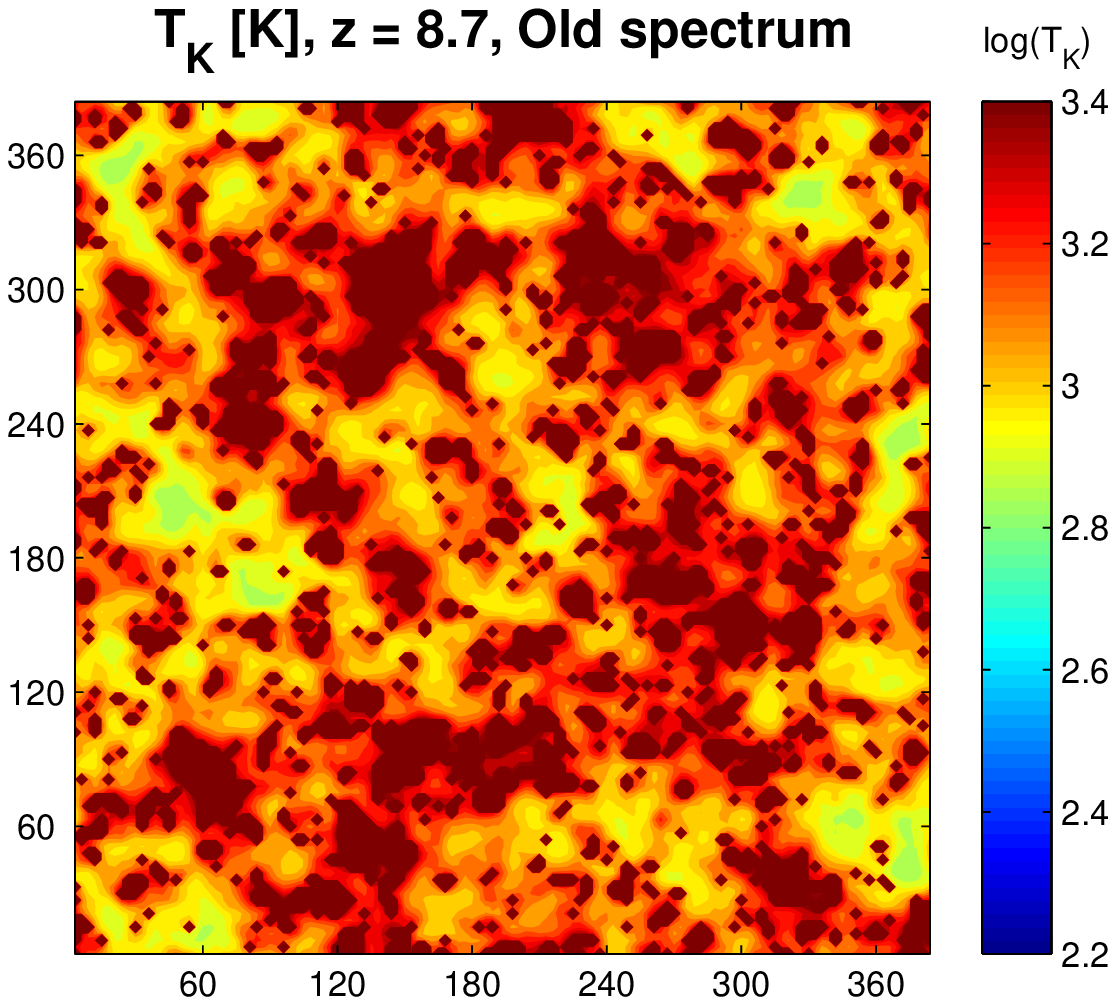}\includegraphics[width=2.6in]{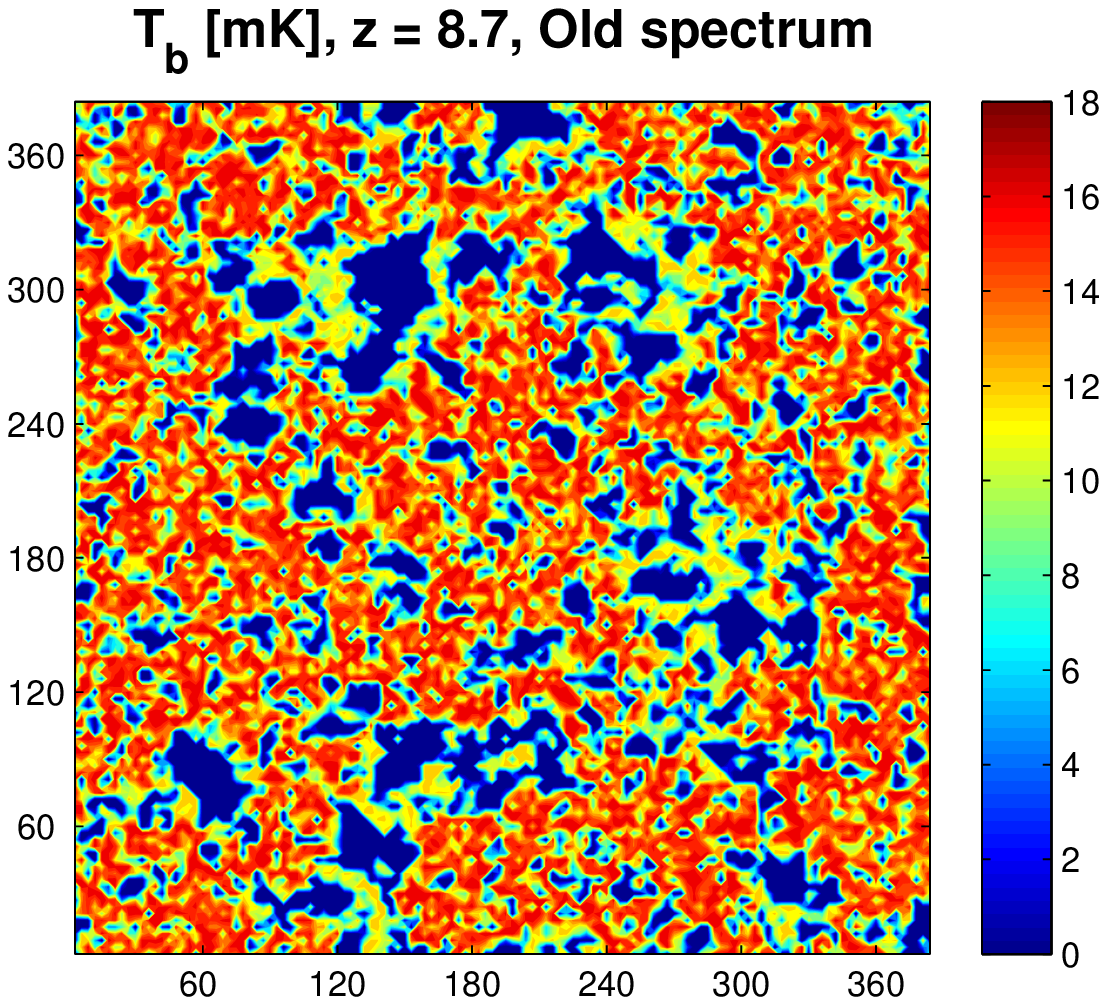}
\includegraphics[width=2.6in]{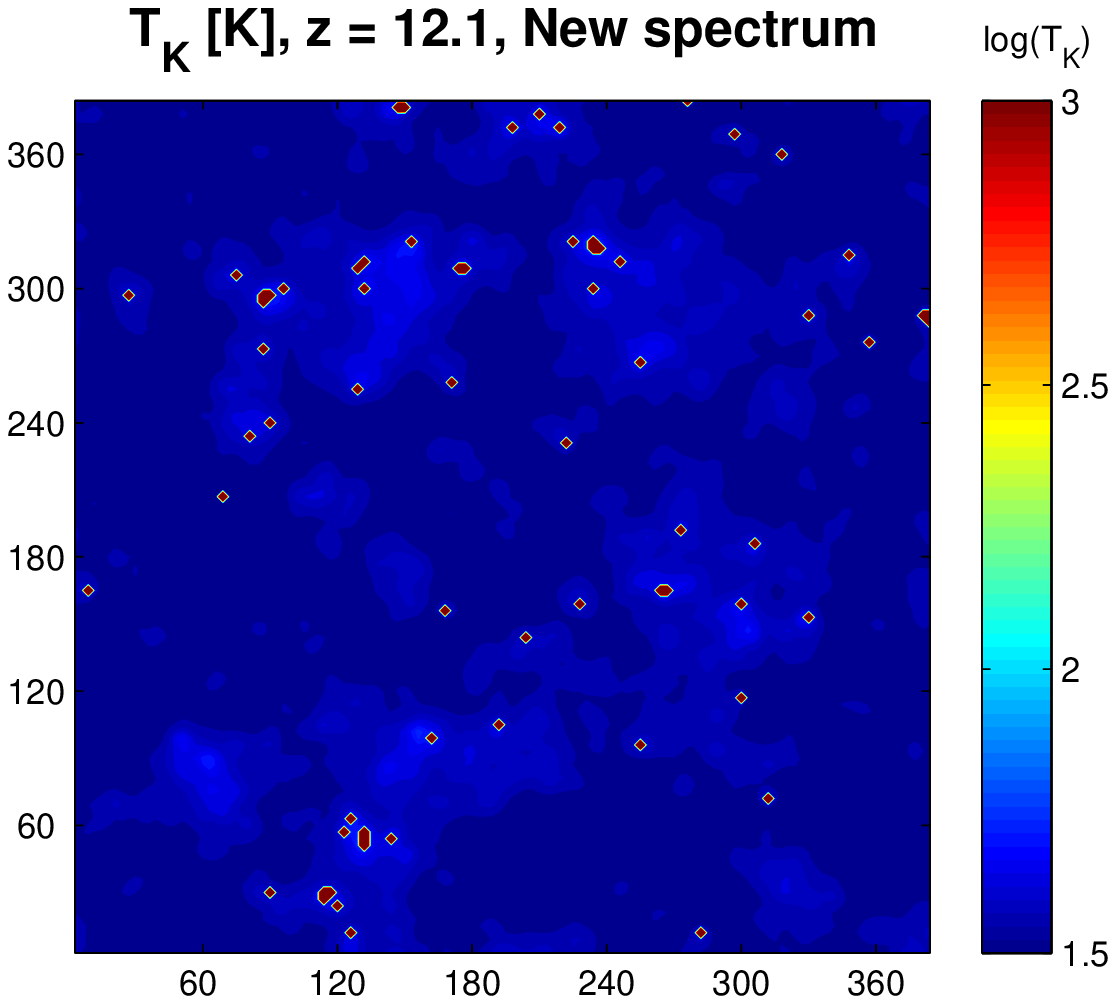}\includegraphics[width=2.6in]{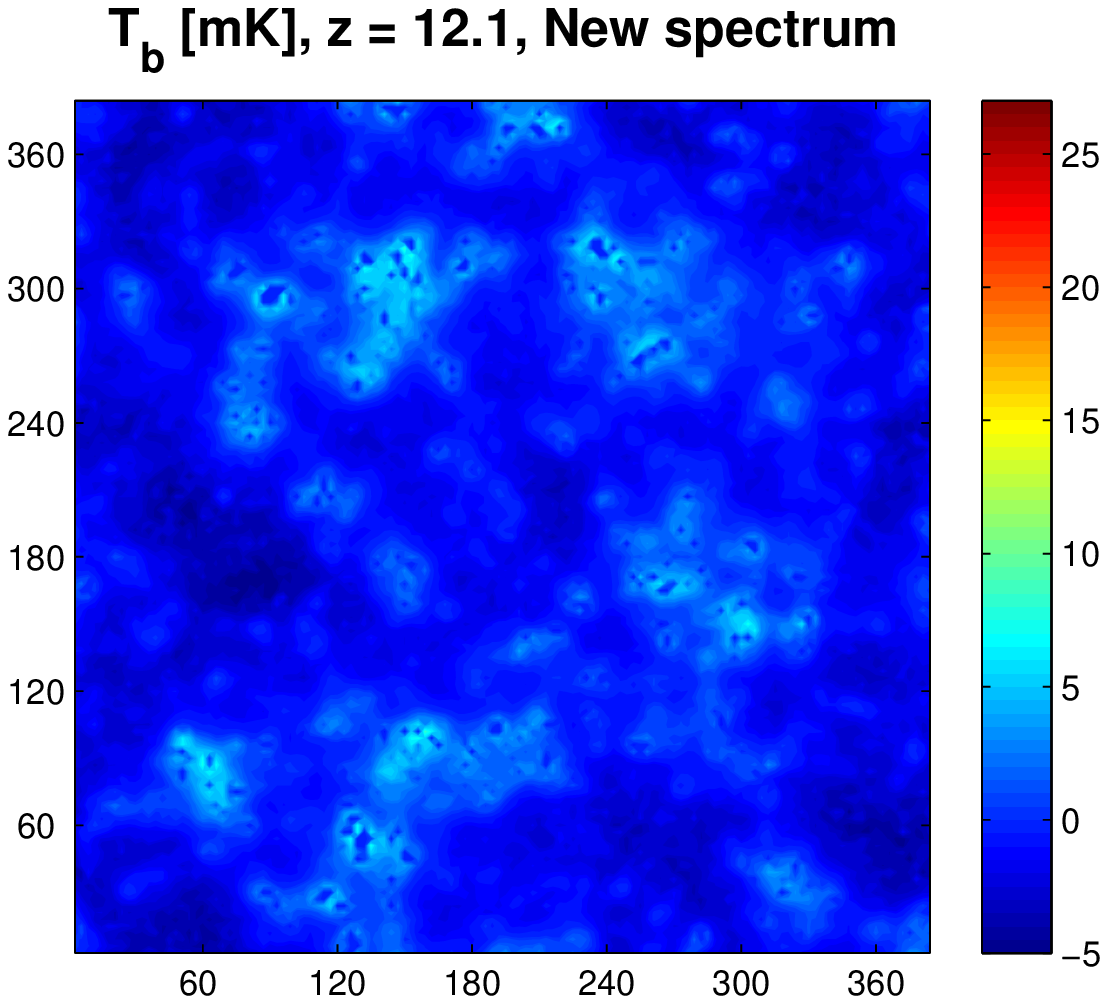}
\includegraphics[width=2.6in]{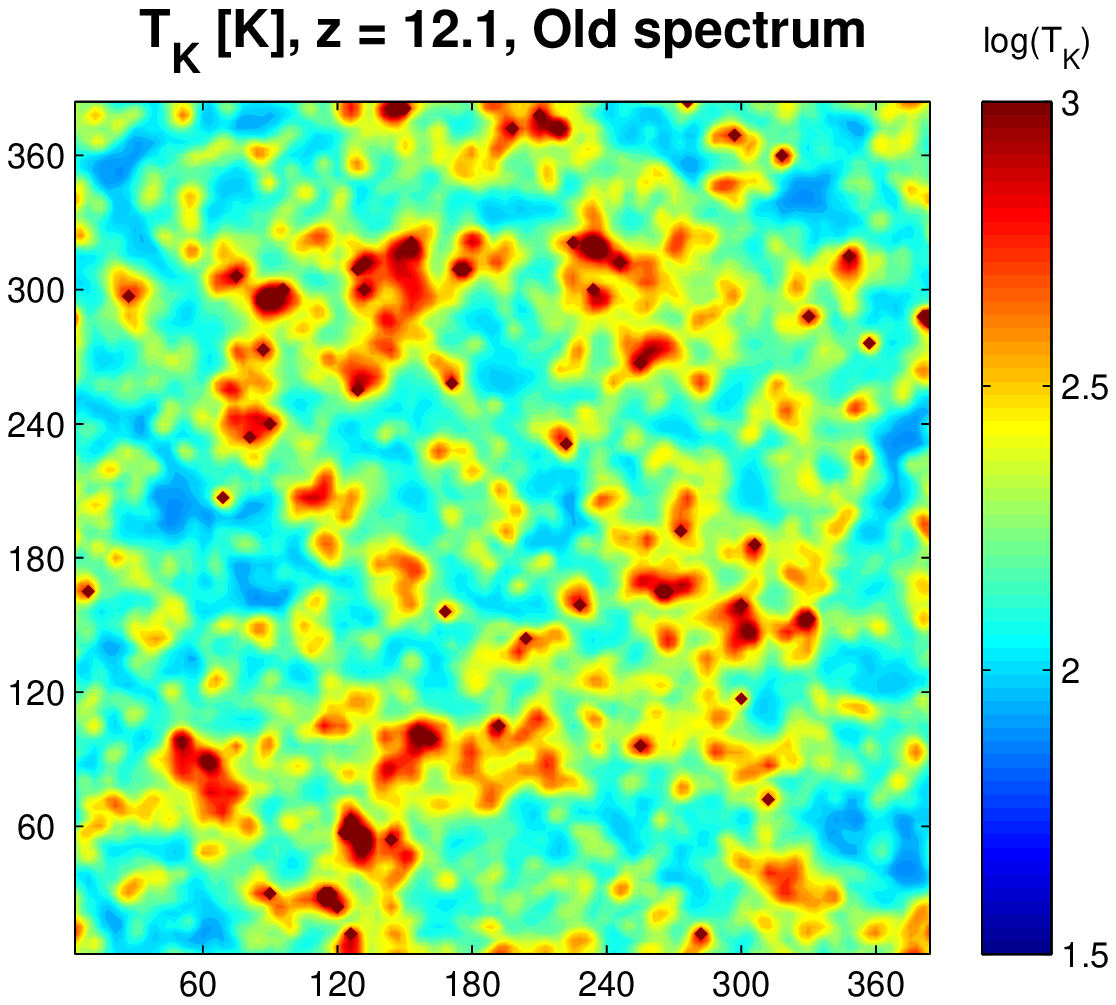}\includegraphics[width=2.6in]{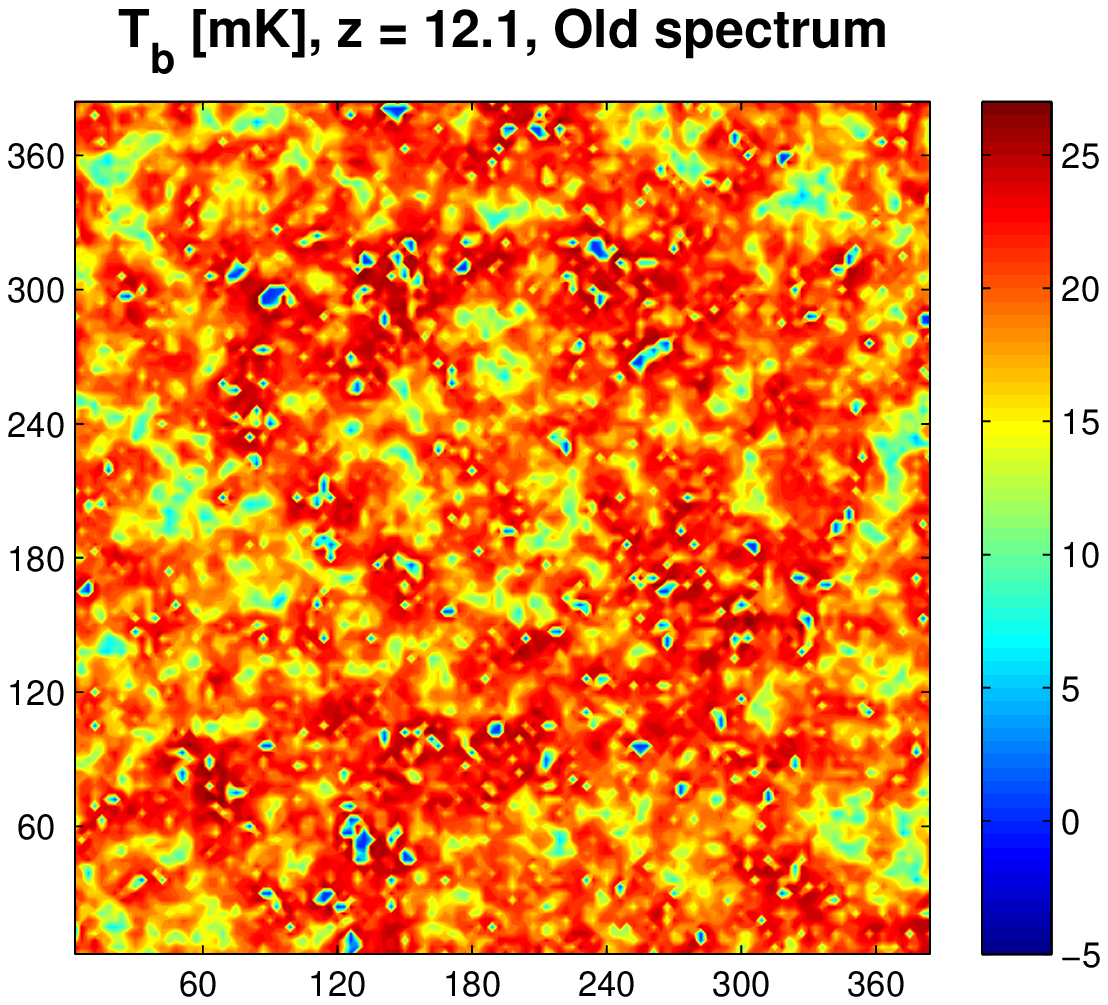}
\caption{Two-dimensional slices of the logarithm of the gas 
temperature $T_K$  in Kelvin (left column) and of the 21-cm brightness
temperature (relative to the CMB) $T_b$ in mK (right column), for the
two cases of atomic cooling halos. Note that the two cases have
identical reionization histories, with fully-ionized pixels shown with
the color corresponding to the hottest $T_K$ (on the left) and to
$T_b=0$ (on the right). From top to bottom: hard X-rays at $z = 8.7$,
which gives ranges $T_K = 165 - 210$~K and $T_b = 2.1 - 14.5$~mK; soft
X-rays at $z = 8.7$, with ranges $T_K = 703-2946$~K and $T_b = 2.4 -
16.8$~mK; hard X-rays at $z = 12.1$, with ranges $T_K = 30.6 - 56.5$~K
and $T_b = -4.8 - 8.4$~mK; and soft X-rays at $z = 12.1$, with ranges
$T_K = 75.9 - 1742$~K and $T_b = 0.8 - 27.4$~mK. Note that $z=12.1$
corresponds to $x_i = 0.14$ and to the heating transition for the 
hard spectrum, while $z=8.7$ corresponds to $x_i = 0.5$. At each
redshift, we use a common scale for the two spectra, for ease
of comparison.}
\label{fig:7}
\end{figure*}

Finally, for visual comparison and physical intuition we show some
snapshots of the gas temperature and the 21-cm brightness temperature
(Figure~\ref{fig:7}). We take two redshifts at which the effect of the
different SEDs is apparent, choosing 8.7 (the midpoint of reionization
for $\tau = 0.075$) and 12.1 (a redshift early in reionization ($x_i =
0.14$), which marks the heating transition for the hard spectrum). We
compare the soft and hard X-ray spectra for the case of atomic
cooling. In this comparison, both cases have the same underlying
distribution of star formation at a given redshift, so they have the
same ionized patches and a similar distribution pattern of gas
temperature and of 21-cm temperature. However, the difference is
visually striking, in that the maps for the hard spectrum are strongly
suppressed both in terms of the absolute values and in the relative
size of the fluctuations.

\section{Summary and Discussion}
\label{Sec:sum}

 The impact of high-mass X-ray binaries on early cosmic heating
  and on the global 21-cm signal has been previously discussed in the
  literature (\S~\ref{Sec:Intro}). However, the case in which X-rays
  emitted by high-redshift sources are inefficient in heating up the
  cosmic gas was considered very extreme, and thus not too
  interesting. In our recent paper \citep{Fialkov:2014} we challenged
  this belief, showing that high-redshift X-ray sources are likely
  inefficient in heating up the Universe and producing fluctuations in
  the gas temperature. While the heating mechanisms at high redshifts
  are still rather unconstrained, in our work we assumed that the
  dominant X-ray sources at high redshifts are HMXBs with hard X-ray
  spectra. This assumption was based on the results of the population
  synthesis simulation by \citet{Fragos:2013}, calibrated to
  low-redshift observations and evolved with redshift accounting for
  the effect of the evolution of metallicity.  Based on current
  knowledge, the contribution of HMXBs to heating likely wins over
  soft X-ray emission from hot gas at high redshifts, given the fact
  that (1) today the total X-ray energy is dominated by X-ray binaries
  over hot gas; (2) both theory and observations suggest that the
  contribution of X-ray binaries increases by an order of magnitude at
  the low metallicities expected at high redshift; and (3) although it
  is true that hard X-rays interact less with the gas, a significant
  fraction of their energy is still absorbed after being redshifted,
  and thus, the fluctuations from soft X-rays are reduced if hard
  X-rays provide a large, uniform background contribution. In this
  paper, in addition to our main "New spectrum" case which we consider
  most likely, we also considered the much softer "Old spectrum".
  Together these two cases reasonably bracket the range of
  possibilities. We believe it is important to consider a range of
  X-ray spectra when making predictions for future 21-cm surveys, since
  in the end only the observations will determine the nature of
  high-redshift heating sources.

Here we discussed for the first time the effect of hard X-rays on the
history of fluctuations in the 21-cm signal during the entire range of
redshifts $z = 7-40$, from the epoch of primordial star formation to
the end of reionization (expanding on \citet{Fialkov:2014b}). This is
a particularly timely as it shows that the signal, which will likely
be observed in near future, may be significantly different than
predicted with the previously-assumed soft X-ray spectrum. The effect
of the hard spectra of HMXBs, which were likely the main source of
X-rays in the early universe \citep{Fragos:2013}, was not considered
in the majority of previous works.

The main consequences of the heating by HMXBs with a hard spectrum
that peaks at $\sim 3$~keV are the following:
\begin{enumerate}
\item A dramatic difference in comparison with soft X-rays is
  that the universe is heated more slowly due to the fact that the
  hard X-rays have longer mean free paths and thus are less
  efficiently absorbed by the cosmic gas. Specifically, the heating
  transition is delayed by a $\Delta z \sim 3$, whereas the variation
  due to the various star-formation scenarios considered here is only
  $\Delta z \sim 0.8$.
\item Since the gas cools adiabatically for much longer, it produces a 
stronger 21-cm absorption signal early on. On the other hand, during
reionization the gas is only moderately warm, and its emission signal
is suppressed. This combination implies that global 21-cm experiments
should focus on $z \sim 20$ rather than on the reionization era.
\item The heating is also much more uniform. As a result, the heating
  fluctuation peak, expected to be found in scenarios with the soft
  X-rays, disappears at intermediate scales of $k >
  0.05$~Mpc$^{-1}$. The wavenumbers at which the heating peak is
  detected in observations should tell us about the characteristic
  mean free path and spectrum of the emitted photons, thus giving key
  clues as to the character of the sources that heated the primordial
  Universe. In addition, the minimum around $z \sim 10 - 12$ which
  separates the heating and ionization domains becomes much
  deeper. The fluctuations are weaker by a factor of $2 - 13$
  (depending on the scale) at the minimum.
\end{enumerate}

The line-of-sight anisotropy makes the analysis of 21-cm fluctuations
much richer, as it in principle allows for three power spectra to be
extracted at each redshift, the isotropic term $\Piso$, its
cross-correlation with density $P_X$, and the density power spectrum
(times a factor of $\bar{T}_b^2$) $P_\delta$. The main conclusions
regarding these components of the power spectra are:
\begin{enumerate}
\item The isotropic term $\Piso$ is the leading contribution to
the total power spectrum at all scales and all the considered epochs.
$P_X$ is sometimes comparable in magnitude but typically smaller by a
factor of a few. $P_\delta$ is the smallest and thus will be the
hardest to measure.
\item $P_X$ is particularly interesting since it changes sign in a
way that encodes information on the various sources of 21-cm
fluctuations. For example, it changes sign at the heating transition,
a moment in which (more generally) the anisotropy of the power
spectrum drops to near zero. Also, $P_X$ is negative when ionization
fluctuations dominate during reionization, and positive at the highest
redshifts. 
\item The dominance of various fluctuation sources depends on 
wavenumber, and can be probed at each redshift from the sign of $P_X$
at various $k$ as well as corresponding slope changes in $\Piso$.
\end{enumerate}

Our predictions affect the expectations for the 21-cm signal in the
range that is observable in the near future. The reionization peak
should be within the sensitivity of present day observatories (such as
LOFAR and the MWA). These experiments may also be able to find signs
of the trough at $z \sim 10-15$, but for hard X-rays the low level of
these fluctuations may require the SKA for detection. At higher
redshifts ($z=15-20$), the expected SKA sensitivity to the power
spectrum ($k^3 P(k)/2\pi^2$) of around a mK$^2$ on large scales
\citep{McQuinn:2006} should allow for a detailed measurement of the
21-cm power spectrum; in particular, a heating peak should be detected
or ruled out. The strong peak of \Lya ~fluctuations at $z\sim 20$ (for
any scenario of star formation considered here) should also be
detectable, since the sensitivity of the SKA is expected to be of
order 10~mK$^2$ at $z\sim 20$. At several of these redshift ranges
(particularly the peaks), the SKA should have sufficient extra
sensitivity to probe the anisotropy of the 21-cm power spectrum and
get a useful measurement of $P_X$. At the same time, global 21-cm
experiments can give complementary information, such as verifying late
heating by measuring a deep minimum in the global signal below
$-140$~mK.

\section{Acknowledgments}
A.F.\ was supported by the LabEx ENS-ICFP:
ANR-10-LABX-0010/ANR-10-IDEX- 0001-02 PSL and NSF grant AST-1312034.
R.B.\ acknowledges Israel Science Foundation grant 823/09 and the
Ministry of Science and Technology, Israel.

\end{document}